\documentclass[12pt,preprint]{aastex}
\shorttitle{Relationship between the GRB pulse width and energy}
\shortauthors{Qin et al.}

\begin{document}

\title{Relationship between the gamma-ray burst pulse width and energy due to the Doppler effect of fireballs}

\author{Y.-P. Qin\altaffilmark{1,2,4}, Y.-M. Dong\altaffilmark{1,3}, R.-J. Lu\altaffilmark{1,2,3}, B.-B. Zhang\altaffilmark{1,3},
 L.-W. Jia\altaffilmark{1,3}}

\altaffiltext{1}{National Astronomical Observatories/Yunnan
Observatory, Chinese Academy of Sciences, P. O. Box 110, Kunming,
Yunnan, 650011, P. R. China}

\altaffiltext{2}{Physics Department, Guangxi University, Nanning,
Guangxi 530004, P. R. China}

\altaffiltext{3}{Graduate School of The Chinese Academy of
Sciences}

\altaffiltext{4}{E-mail:ypqin@ynao.ac.cn}

\begin{abstract}
We study in details how the pulse width of gamma-ray bursts is
related with energy under the assumption that the sources
concerned are in the stage of fireballs. Due to the Doppler effect
of fireballs, there exists a power law relationship between the
two quantities within a limited range of frequency. The power law
range and the power law index depend strongly on the observed peak
energy $E_p$ as well as the rest frame radiation form, and the
upper and lower limits of the power law range can be determined by
$E_p$. It is found that, within the same power law range, the
ratio of the $FWHM$ of the rising portion to that of the decaying
phase of the pulses is also related with energy in the form of
power laws. A platform-power-law-platform feature could be
observed in the two relationships. In the case of an obvious
softening of the rest frame spectrum, the two power law
relationships also exist, but the feature would evolve to a peaked
one. Predictions on the relationships in the energy range covering
both the BATSE and Swift bands for a typical hard burst and a
typical soft one are made. A sample of FRED (fast rise and
exponential decay) pulse bursts shows that 27 out of the 28
sources belong to either the platform-power-law-platform feature
class or the peaked feature group, suggesting that the effect
concerned is indeed important for most of the sources of the
sample. Among these bursts, many might undergo an obvious
softening evolution of the rest frame spectrum.
\end{abstract}


\section{Introduction}

Owing to the large amount of energy observed, gamma-ray bursts
(GRBs) were assumed to undergo a stage of fireballs which expand
relativistically (see, e.g., Goodman 1986; Paczynski 1986).
Relativistic bulk motion of the gamma-ray-emitting plasma would
play a role in producing the observed phenomena of the sources
(Krolik \& Pier 1991). It was believed that the Doppler effect
over the whole fireball surface (the so-called relativistic
curvature effect) might be the key factor to account for the
observed spectrum of the events (see, e.g., Meszaros \& Rees 1998;
Hailey et al. 1999; Qin 2002, 2003).

Some simple bursts with well-separated structure suggest that they
may consist of fundamental units of radiation such as pulses, with
some of them being seen to comprise a fast rise and an exponential
decay (FRED) phases (see, e.g., Fishman et al. 1994). These FRED
pulses could be well represented by flexible empirical or
quasi-empirical functions (see e.g., Norris et al. 1996; Kocevski
et al. 2003). Fitting the corresponding light curves with the
empirical functions, many statistical properties of pulses were
revealed. Light curves of GRB pulses were found to become narrower
at higher energies (Fishman et al. 1992; Link, Epstein, \&
Priedhorsky 1993). Fenimore et al. (1995) showed that the average
pulse width is related with energy by a power law with an index of
about $-0.4$. This was confirmed by later studies (Fenimore et al.
1995; Norris et al. 1996, 2000; Costa 1998; Piro et al. 1998;
Nemiroff 2000; Feroci et al. 2001; Crew et al. 2003).

In the past few years, many attempts of interpretation of light
curves of GRBs have been made (see, e.g., Fenimore et al. 1996;
Norris et al. 1996; Norris et al. 2000; Ryde \& Petrosian 2002;
Kocevski et al. 2003). It was suggested that the power law
relationship could be attributed to synchrotron radiation (see
Fenimore et al. 1995; Cohen et al. 1997; Piran 1999). Kazanas,
Titarchuk, \& Hua (1998) proposed that the relationship could be
accounted for by synchrotron cooling (see also Chiang 1998; Dermer
1998; and Wang et al. 2000). Phenomena such as the
hardness-intensity correlation and the FRED form of pulses were
recently interpreted as signatures of the relativistic curvature
effect (Fenimore et al. 1996; Ryde \& Petrosian 2002; Kocevski et
al. 2003; Qin et al. 2004, hereafter Paper I). It was suspected
that the power law relationship might result from a relative
projected speed or a relative beaming angle (Nemiroff 2000). Due
to the feature of self-similarity across energy bands observed
(see, e.g., Norris et al. 1996), it is likely that the observed
difference between different channel light curves might mainly be
due to the energy channels themselves. In other words, light
curves of different energy channels might arise from the same
mechanism (e.g., parameters of the rest frame spectrum and
parameters of the expanding fireballs are the same for different
energy ranges), differing only in the energy ranges involved.

We believe that, if different channel light curves of a burst can
be accounted for by the same mechanism where, except the energy
ranges concerned, no parameters are allowed to be different for
different energy channels, then the mechanism must be the main
cause of the observed difference. A natural mechanism that
possesses this property might be the Doppler effect of the
expanding fireball surface when a rest frame radiation form is
assumed. Indeed, as shown in Paper I, the four channel light
curves of GRB 951019 were found to be well fitted by a single
formula derived when this effect was taken into account. In Paper
I, the power law relationship between the pulse width and energy
was interpreted as being mainly due to different active areas of
the fireball surface corresponding to the majority of photons of
these channels. However, how the width is related with energy
remains unclear. Presented in the following is a detailed analysis
on this issue and based on it predictions on the relationship over
a wide band covering those of BATSE and Swift will be made.

This paper is organized as follows. In section 2, we investigate
in a general manner how the width and the ratio of the rising
width to the decaying width of GRB pulses are related with energy.
Then we make predictions on the relationship over the BATSE and
Swift bands for a typical hard and a typical soft bursts in
section 3. In section 4, a sample containing 28 FRED pulse sources
is employed to illustrate the relationship. A brief discussion and
conclusions are presented in the last section.

\section{General analysis on the relationship}

Studies of the Doppler effect of the expanding fireball surface
were presented by different authors, and based on these studies
formulas applicable to various situations are available (see,
e.g., Fenimore et al. 1996; Granot et al. 1999; Eriksen \& Gron
2000; Dado et al. 2002a, 2002b; Ryde \& Petrosian 2002; Kocevski
et al. 2003; Paper I; Shen et al. 2005). Employed in the following
is one of them suitable for studying the issue concerned above,
where a highly symmetric and expanding fireball is concerned.

It can be verified that the expected flux of a fireball expanding
with a Lorentz factor $\Gamma >1$ can be determined by (for a
detailed derivation for this form of formula one can refer to
Paper I)
\begin{eqnarray}
f_\nu (\tau )= \frac{2\pi R_c^2}{D^2\Gamma ^3(1-\beta )^2(1+\frac
\beta
{1-\beta }\tau )^2}\nonumber\\
\times\int_{\widetilde{\tau }_{\theta ,\min }}^{\widetilde{\tau
}_{\theta ,\max }}\widetilde{I}(\tau _\theta )(1+\beta \tau
_\theta )^2(1-\tau +\tau _\theta )g_{0,\nu }(\nu _{0,\theta
})d\tau _\theta ,
\end{eqnarray}
with $\tau _{\min }\leq \tau \leq \tau _{\max }$, $\tau _{\min
}\equiv (1-\beta )\tau _{\theta ,\min }$, $\tau _{\max }\equiv
1+\tau _{\theta ,\max }$, $\tau \equiv (t-\frac
Dc+\frac{R_c}c-t_c)/\frac{R_c}c$, and $\tau _\theta \equiv
(t_\theta -t_c)/\frac{R_c}c$, where $t$ is the observation time
measured by the distant observer, $t_\theta $ is the local time
measured by the local observer located at the place encountering
the expanding fireball surface at the position of $\theta $
relative to the center of the fireball, $t_c$ is the initial local
time, $R_c$ is the radius of the fireball measured at $t_\theta
=t_c$, $D$ is the distance from the fireball to the observer,
$\widetilde{I}(\tau _\theta )$ represents the development of the
intensity measured by the local observer, and $g_{0,\nu }(\nu
_{0,\theta })$ describes the rest frame radiation, and $\nu
_{0,\theta }=(1-\beta +\beta \tau )\Gamma \nu /(1+\beta \tau
_\theta )$, $\widetilde{ \tau }_{\theta ,\min }=\max \{\tau
-1,\tau _{\theta ,\min }\}$, and $ \widetilde{\tau }_{\theta ,\max
}=\min \{\tau /(1-\beta ),\tau _{\theta ,\max }\}$, with $\tau
_{\theta ,\min} = (t_{\theta ,\min} -t_c)/\frac{R_c}c$ and $\tau
_{\theta ,\max} = (t_{\theta ,\max} -t_c)/\frac{R_c}c$ being the
upper and lower limits of $\tau _\theta $ confining
$\widetilde{I}(\tau _\theta )$, respectively. (Note that, since
the limit of the Lorentz factor is $\Gamma >1$, the formula can be
applied to the cases of relativistic, sub-relativistic, and
non-relativistic motions.)

The expected count rate of the fireball measured within frequency
interval $ [\nu _1,\nu _2]$ can be calculated with
\begin{eqnarray}
C(\tau )=\int_{\nu _1}^{\nu _2}\frac{f_\nu (\tau )}{h\nu }d\nu =
\frac{2\pi R_c^2}{hD^2\Gamma ^3(1-\beta )^2(1+\frac \beta {1-\beta
}\tau )^2}\nonumber\\
\times\int_{\widetilde{\tau }_{\theta ,\min }}^{\widetilde{\tau }
_{\theta ,\max }}[\widetilde{I}(\tau _\theta )(1+\beta \tau
_\theta )^2(1-\tau +\tau _\theta )\int_{\nu _1}^{\nu
_2}\frac{g_{0,\nu }(\nu _{0,\theta })}\nu d\nu ]d\tau _\theta .
\end{eqnarray}
It suggests that, except the mechanism [i.e., $\widetilde{I}(\tau
_\theta )$ and $g_{0,\nu }(\nu _{0,\theta })$] and the state of
the fireball (i.e., $ \Gamma $, $R_c$ and $D$), light curves of
the source depend on the energy range as well.

For the sake of simplicity, we first employ a local pulse to study
the relationship in a much detail and later employ other local
pulses to study the same issue in less details. The local pulse
considered in this section is that of Gaussian which is assumed to
be
\begin{equation}
\widetilde{I}(\tau _\theta )= \\
I_0\exp [-(\frac{\tau _\theta -\tau _{\theta ,0} }\sigma
)^2]\qquad \qquad \qquad (\tau _{\theta ,\min }\leq \tau _\theta
),
\end{equation}
where $I_0$, $\sigma $, $\tau _{\theta ,0}$ and $\tau _{\theta
,\min }$ are constants. As shown in Paper I, there is a constraint
to the lower limit of $ \tau _\theta $, which is $\tau _{\theta
,\min }>-1/\beta $. Due to this constraint, it is impossible to
take a negative infinity value of $\tau _{\theta ,\min }$ and
therefore the interval between $\tau _{\theta ,0}$ and $\tau
_{\theta ,\min }$ must be limited. Here we assign $\tau _{\theta
,0}=10\sigma +\tau _{\theta ,\min }$ so that the interval between
$\tau _{\theta ,0}$ and $\tau _{\theta ,\min }$ would be large
enough to make the rising part of the local pulse close to that of
the Gaussian pulse. The $ FWHM $ of the Gaussian pulse is $\Delta
\tau _{\theta ,FWHM}=2\sqrt{ \ln 2}\sigma $, which leads to
$\sigma =\Delta \tau _{\theta ,FWHM}/2\sqrt{ \ln 2}$. In the
following, we assign $\tau _{\theta ,\min }=0$, and take $ \Delta
\tau _{\theta ,FWHM}=0.01$, $0.1$, $1$, and $10$, and adopt
$\Gamma =10$, $100$, and $1000$.

\subsection{The case of a typical Band function}

Here we employ the Band function (Band et al. 1993) with typical
indexes $\alpha _0=-1$ and $\beta _0=-2.25$ as the rest frame
radiation form to investigate how the $FWHM$ and the
$FWHM1$/$FWHM2$ are related with the corresponding energy, where
$FWHM1$ and $FWHM2$ are the $FWHM$ in the rising and decaying
phases of the light curve, respectively. The $FWHM$ and
$FWHM1$/$FWHM2$ of the observed light curve arising from the local
Gaussian pulse associated with certain frequency could be well
determined according to (2), when (3) is applied. Displayed in
Figs. 1a and 1b are the $ FWHM$--- $\nu /\nu _{0,p}$ and
$FWHM1/FWHM2$--- $\nu /\nu _{0,p}$ curves, respectively. One finds
from these curves that, for all sets of the parameters adopted
here, a semi-power law relationship between each of the two
quantities ($FWHM$ and $FWHM1$/$FWHM2$) and $\nu /\nu _{0,p}$
could be observed within a range (called the power law range)
spanning over more than 1 order of magnitudes of frequency. Beyond
this range (i.e., in higher and lower frequency bands), both the
$FWHM$ and $FWHM1/FWHM2$ of the observed light curve would remain
unchanged with frequency. We call the unchanged section of the
curves in lower frequency band relative to the power law range a
lower band platform and call that in higher frequency band a
higher band platform. For a certain rest frame spectrum (say, when
the value of $\nu _{0,p}$ is fixed), the power law range shifts to
higher energy bands when $\Gamma $ becomes larger. The power law
range could therefore become an indicator of the Lorentz factor as
long as $\nu _{0,p}$ is fixed (in practice, as $\nu _{0,p}$ is
always unclear, what can be determined is the product $\Gamma \nu
_{0,p}$ which is directly associated with the observed peak energy
$E _{p}$; see what discussed below).

The power law range shown in a $ FWHM$--- $\nu /\nu _{0,p}$ curve
is marked by a smooth turning at its lower energy end and a sharp
turning at its higher end. Let $\nu _{low}$ (or $E_{low}$) denote
the position of the turning at the lower energy end and $\nu
_{high}$ (or $E_{high}$) represent that at the higher end. One
finds that $\nu _{high} $ would be well defined due to the sharp
feature associated with it while $\nu _{low}$ would not since the
corresponding feature is smooth. According to Fig. 1a, we simply
define $\nu _{low}$ by $\log FWHM(\nu _{low})\equiv \log
FWHM_{\max }-(\log FWHM_{\max }-\log FWHM_{\min })/10$, where
$FWHM_{\min }$ and $FWHM_{\max }$ are the minimum and maximum
values of the $FWHM$ of the light curves. Listed in Table 1 are
the values of $\nu _{low}$ and $\nu _{high}$ as well as
$FWHM_{\min }$ and $FWHM_{\max }$ deduced from the curves of Fig.
1a. One can conclude from this table that $\nu _{high}\simeq
2.4\sim 2.5\Gamma \nu _{0,p}$ for all the adopted Lorentz factors
($\Gamma =10$, $100$ and $1000$), and $\log \nu _{high}-\log \nu
_{low}\simeq 1.19\sim 1.26$, $1.20\sim 1.25$ and $1.20\sim 1.26$,
for $\Gamma =10$, $100$ and $1000$, respectively. It reveals that
$\nu _{high}$ is proportional to $\Gamma \nu _{0,p}$. For the same
value of $\Gamma \nu _{0,p}$, $\nu _{high}$ would be independent
of $\Gamma $ or $\nu _{0,p}$. The power law range spans over more
than 1 order of magnitudes of frequency for all the Lorentz
factors concerned. In addition, we find that, for the same value
of $\Delta \tau _{\theta ,FWHM}$, $FWHM_{\min }\propto \Gamma
^{-2}$ and $FWHM_{\max }\propto \Gamma ^{-2}$.

As shown in Qin (2002), when taking into account the Doppler
effect of fireballs, the observed peak frequency would be related
with the peak frequency of the typical rest frame Band function
spectrum by $\nu_{p} \simeq 1.67\Gamma\nu_{0,p}$, i.e.,
$E_{p}\simeq 1.67\Gamma E_{0,p}$. In terms of $E_{p}$, we get from
Table 1 that $log E_{low}- log E_{p} \simeq (-1.10 \sim -1.02)$
and $log E_{high}- log E_{p} \simeq (0.157 \sim 0.177)$.

In the same way, we confine the power law range shown in Fig. 1b
with $\nu _{low}$ and $\nu _{high}$ as well, with $\nu _{low}$
being defined by $\log (FWHM1/FWHM2)(\nu _{low})\equiv \log
(FWHM1/FWHM2)_{\min }+[\log (FWHM1/FWHM2)_{\max }-\log
(FWHM1/FWHM2)_{\min }]/10$. Listed in Table 2 are the values of
$\nu _{low}$, $\nu _{high}$, $(FWHM1/FWHM2)_{\min }$ and $
(FWHM1/FWHM2)_{\max }$ obtained from the curves of Fig. 1b. We
find from this table that $\nu _{high}\simeq 2.3\sim 2.4\Gamma \nu
_{0,p}$, $ 2.3\sim 2.5\Gamma \nu _{0,p}$ and $2.2\sim 2.5\Gamma
\nu _{0,p}$, for $ \Gamma =10$, $100$ and $1000$, respectively,
and $\log \nu _{high}-\log \nu _{low}\simeq 1.18\sim 1.25$,
$1.20\sim 1.24$ and $1.20\sim 1.25$, for $ \Gamma =10$, $100$ and
$1000$, respectively. It also shows that, although both $\nu
_{high}$ and $\nu _{low}$ are proportional to $\Gamma \nu _{0,p}$,
they are independent of $\Gamma $ or $\nu _{0,p}$ alone. In terms
of $E_{p}$, we get $log E_{low}- log E_{p} \simeq (-1.10 \sim
-1.02)$ and $log E_{high}- log E_{p} \simeq (0.127 \sim 0.167)$.
In addition, we find that, for the same value of $\Delta \tau
_{\theta ,FWHM}$, $FWHM_{\min }\propto \Gamma ^{-2}$ and
$FWHM_{\max }\propto \Gamma ^{-2}$. Suggested by Table 2, the
values of $ (FWHM1/FWHM2)_{\min }$ and $(FWHM1/FWHM2)_{\max }$
rely only on the local pulse width $\Delta \tau _{\theta ,FWHM}$.

The relation between $E_{low}$ and $E_{p}$ or that between
$E_{high}$ and $ E_{p}$ suggests that once we obtain the value of
$E_{low}$ or $E_{high}$ in the case of the typical rest frame Band
function spectrum, we would be able to estimate $ E_{p}$, or vice
versa.

It is noticed that a certain value of $\nu /\nu _{0,p}$ might
correspond to different energies associated with different values
of $ \nu _{0,p}$. Let us assign $\nu _{0,p}=10keVh^{-1}$ when
taking $\Gamma =10$ , $\nu _{0,p}=1keVh^{-1}$ when taking $\Gamma
=100$, and $\nu _{0,p}=0.1keVh^{-1}$ when taking $\Gamma =1000$.
In this situation, $\Gamma \nu _{0,p}=100keVh^{-1}$ holds for all
these cases. Presented in Fig. 1c are the curves of Fig. 1a in
terms of energy, where the power law range confined by $\log
E_{low}/keV=1.13$ and $\log E_{high}/keV=2.38$ (see Table 1) is
displayed. Shown in Fig. 1d are the curves of Fig. 1b in terms of
energy, where the power law range confined by $\log
E_{low}/keV=1.12$ and $ \log E_{high}/keV=2.36$ (see Table 2) is
plotted. From Fig. 1d, one finds that, the curves corresponding to
$\Gamma =10$, $100$ and $1000$ are hard to be distinguishable.
When $\Gamma \nu _{0,p}$ being the same (here, $\Gamma \nu
_{0,p}=100keVh^{-1}$), the two relationships (one is that between
$FWHM$ and energy, and the other is that between $FWHM1/FWHM2$ and
energy) are independent of the Lorentz factor, and the power law
ranges of the curves arising from $\Gamma =10$, $100$ and $1000$
become almost the same.

One can conclude from this analysis that, in the case of adopting
the typical Band function with $\alpha _0=-1$ and $\beta _0=-2.25$
as the rest frame radiation form, there exists a semi-power law
relationship spanning over more than one order of magnitudes of
energy, between the width of pulses and energy as well as between
the ratio of the rising width to the decaying width of pulses and
energy. The upper and lower limits of this power law range are
well related with the observed peak energy $E_p$ of a fireball
source.

\subsection{The case of other spectra}

Adopted as the rest frame radiation form, let us consider two
other spectra which are much different from the Band function
(especially in the high energy band). One is the thermal
synchrotron spectrum: $I_\nu \propto (\nu /\nu _{0,s})\exp [-(\nu
/\nu _{0,s})^{1/3}]$, where $\nu _{0,s}$ is a constant including
all constants in the exponential index (Liang et al. 1983). The
other is the Comptonized spectrum: $I_\nu \propto \nu ^{1+\alpha
_{0,C}}\exp (-\nu /\nu _{0,C})$, where $\alpha _{0,C}$ and $\nu
_{0,C}$ are constants. Typical value $\alpha _{0,C}=-0.6$
(Schaefer et al. 1994) for the index of the Comptonized radiation
will be adopted.

Presented in Figs. 2a and 2b are the $FWHM$ --- $\nu /\nu _{0,s}$
and $ FWHM1/FWHM2$ --- $\nu /\nu _{0,s}$ curves, respectively,
corresponding to the rest frame thermal synchrotron spectrum and
local Gaussian pulse (3). A semi-power law relationship could also
be observed in both plots. In the case of $\Delta \tau _{\theta
,FWHM}=1$ (where the turn over could be well defined in the two
plots) we get from Fig. 2a that $\nu _{high}\simeq 1.6\times
10^5\Gamma \nu _{0,s}$, $1.7\times 10^5\Gamma \nu _{0,s}$ and
$1.7\times 10^5\Gamma \nu _{0,s}$, for $\Gamma =10$, $100$ and
$1000$, respectively, and $\log \nu _{high}-\log \nu _{low}\simeq
5.30$, $5.53$ and $5.66$, for $\Gamma =10$, $ 100$ and $1000$,
respectively, and obtain from Fig. 2b that $\nu _{high}\simeq
4.6\times 10^5\Gamma \nu _{0,s}$, $4.8\times 10^5\Gamma \nu
_{0,s}$ and $4.7\times 10^5\Gamma \nu _{0,s}$, for $\Gamma =10$,
$100$ and $ 1000$, respectively, and $\log \nu _{high}-\log \nu
_{low}\simeq 5.75$, $ 5.99 $ and $6.09$, for $\Gamma =10$, $100$
and $1000$, respectively. It suggests that, in the case of the
rest frame thermal synchrotron spectrum, $ \nu _{high}$ is
proportional to $\Gamma \nu _{0,s}$, and the power law range can
span over more than 5 orders of magnitudes of frequency.

Shown in Figs. 2c and 2d are the $FWHM$ --- $\nu /\nu _{0,p}$ and
$ FWHM1/FWHM2$ --- $\nu /\nu _{0,p}$ curves, respectively,
associated with the rest frame Comptonized spectrum and arising
from local Gaussian pulse (3). A semi-power law relationship could
also be detected in both plots. In the case of $\Delta \tau
_{\theta ,FWHM}=1 $ we deduce from Fig. 2c that $\nu _{high}\simeq
1.2\times 10^2\Gamma \nu _{0,C}$, $1.2\times 10^2\Gamma \nu
_{0,C}$ and $1.3\times 10^2\Gamma \nu _{0,C}$, for $\Gamma =10$,
$100$ and $1000$, respectively, and $\log \nu _{high}-\log \nu
_{low}\simeq 2.44$, $2.45$ and $2.46$, for $\Gamma =10$, $ 100$
and $1000$, respectively, and find from Fig. 2d that $\nu
_{high}\simeq 1.2\times 10^2\Gamma \nu _{0,C}$, $1.1\times
10^2\Gamma \nu _{0,C}$ and $ 1.2\times 10^2\Gamma \nu _{0,C}$, for
$\Gamma =10$, $100$ and $1000$, respectively, and $\log \nu
_{high}-\log \nu _{low}\simeq 2.45$, $2.43$ and $ 2.45$, for
$\Gamma =10$, $100$ and $1000$, respectively. In the case of the
rest frame Comptonized spectrum, $\nu _{high}$ is proportional to
$\Gamma \nu _{0,C}$. The power law range spans over more than 2
orders of magnitudes of frequency.

It could be concluded that, in the case of adopting a rest frame
spectrum with an exponential tail in the high energy band, a
semi-power law relationship between the $FWHM$ and energy or
between $ FWHM1/FWHM2$ and energy could also be observed. The
range (spanning over more than 2 orders of magnitudes of energy)
would be much larger than that in the case of the Band function.
It seems common that, for a rest frame spectrum, there exists a
power law relationship between each of the $FWHM$ and $
FWHM1/FWHM2$ and energy within an energy range. The range is very
sensitive to the rest frame spectrum and the product of the rest
frame peak energy and the Lorentz factor.

\subsection{The case of the rest frame radiation form
varying with time}

It has been known that indexes of spectra of many GRBs are
observed to vary with time (see Preece et al. 2000). We are
curious how the relationship would be if the rest frame spectrum
develops with time. Here, corresponding to the soft-to-hard
phenomenon, let us consider a simple case where the rest frame
spectrum is a Band function with its indexes and the peak
frequency decreasing with time. We assume a simple evolution of
indexes $\alpha _0$ and $\beta _0$ and peak frequency $\nu_{0,p}$
which follow $\alpha _0=-0.5-k(\tau _\theta -\tau _{\theta ,1
})/(\tau _{\theta ,2 }-\tau _{\theta ,1 })$, $\beta _0=-2-k(\tau
_\theta -\tau _{\theta ,1 })/(\tau _{\theta ,2 }-\tau _{\theta ,1
})$ and $log \nu_{0,p}=0.1-k(\tau _\theta -\tau _{\theta ,1
})/(\tau _{\theta ,2 }-\tau _{\theta ,1 })$, for $\tau _{\theta ,1
}\leq \tau _{\theta}\leq \tau _{\theta ,2 }$. For $\tau _{\theta}
< \tau _{\theta ,1 }$, $\alpha _0=-0.5$, $\beta _0=-2$ and $log
\nu_{0,p}=0.1$, while for $\tau _{\theta} > \tau _{\theta ,2 }$,
$\alpha _0=-0.5-k$, $\beta _0=-2-k$ and $log \nu_{0,p}=0.1-k$. We
take $k=0.1$, $0.5$ and $1$ respectively (they correspond to
different speeds of decreasing) and adopt $\Gamma =10$, $100$,
$1000$ respectively in the following analysis.

Let us employ local Gaussian pulse (3) with $\Delta \tau _{\theta
,FWHM}=0.1$ to study the relationship. We adopt $\tau _{\theta
,1}=9\sigma +\tau _{\theta ,\min }$ and $\tau _{\theta
,2}=11\sigma +\tau _{\theta ,\min }$ and once more assign $\tau
_{\theta ,0}=10\sigma +\tau _{\theta ,\min }$ and $\tau _{\theta
,\min }=0$ (see what mentioned above). Displayed in Fig. 3 are the
expected $FWHM$ --- $\nu /\nu _{0,p,max}$ and $FWHM1/FWHM2$
--- $\nu /\nu _{0,p,max}$ curves, where the frequency is presented
in units of $\nu_{0,p,max}$ which is the largest value of
$\nu_{0,p}$ adopted.

We find that, when the decreasing speed becomes larger (say,
$k=0.5$ or $1$), the relationships would obviously betray what
noticed above. In this situation, the relationship between the
pulse width and energy could show at least two semi-power law
ranges with the index of that in the lower energy band being
positive, and therefore a peak value of the width marking the two
lower energy power law ranges would be observed. Accordingly, the
lower band platform noticed above disappears. In the case of the
relationship between $FWHM1/FWHM2$ and $\nu$, a peak of
$FWHM1/FWHM2$ marking two higher energy semi-power law ranges
could also be detected. This peaked feature is a remarkable
signature of the evolution of the rest frame spectrum.

\subsection{The case of other local pulses}

Here we investigate if different local pulses would lead to a much
different result. Three forms of local power law pulses are
considered. We choose power law forms instead of other local pulse
forms due to the fact that different values of the power law index
would correspond to entirely different forms of local pulses.

The first is the local pulse with a power law rise and a power law
decay, which is assumed to be
\begin{equation}
\widetilde{I}(\tau _\theta )=I_0\{
\begin{array}{c}
(\frac{\tau _\theta -\tau _{\theta ,\min }}{\tau _{\theta ,0}-\tau
_{\theta ,\min }})^\mu \qquad \qquad \qquad \qquad \qquad (\tau
_{\theta ,\min }\leq
\tau _\theta \leq \tau _{\theta ,0}) \\
(1-\frac{\tau _\theta -\tau _{\theta ,0}}{\tau _{\theta ,\max
}-\tau _{\theta ,0}})^\mu \qquad \qquad \qquad \qquad (\tau
_{\theta ,0}<\tau _\theta \leq \tau _{\theta ,\max })
\end{array}
,
\end{equation}
where $I_0$, $\mu $, $\tau _{\theta ,\min }$, $\tau _{\theta ,0}$
and $\tau _{\theta ,\max }$ are constants. The peak of this
intensity is at $\tau _{\theta ,0}$, and the two $FWHM$ positions
of this intensity before and after $\tau _{\theta ,0}$ are $\tau
_{\theta ,FWHM1}=2^{-1/\mu }\tau _{\theta ,0}+(1-2^{-1/\mu })\tau
_{\theta ,\min }$ and $\tau _{\theta ,FWHM2}=2^{-1/\mu }\tau
_{\theta ,0}+(1-2^{-1/\mu })\tau _{\theta ,\max }$, respectively.
In the case of $\mu =2$, the $FWHM$ of this local pulse is $\Delta
\tau _{\theta ,FWHM}=(1-1/\sqrt{2})(\tau _{\theta ,\max }-\tau
_{\theta ,\min })$, which leads to $\tau _{\theta ,\max }=\Delta
\tau _{\theta ,FWHM}/(1-1/\sqrt{2})+\tau _{\theta ,\min }$. The
second is the local pulse with a power law rise which is written
as
\begin{equation}
\widetilde{I}(\tau _\theta )=I_0(\frac{\tau _\theta -\tau _{\theta
,\min }}{ \tau _{\theta ,\max }-\tau _{\theta ,\min }})^\mu \qquad
\qquad \qquad \qquad \qquad (\tau _{\theta ,\min }\leq \tau
_\theta \leq \tau _{\theta ,\max }).
\end{equation}
The peak of this intensity is at $\tau _{\theta ,\max }$. In the
case of $ \mu =2$, the relation of $\tau _{\theta ,\max }=\Delta
\tau _{\theta ,FWHM}/(1-1/\sqrt{2})+\tau _{\theta ,\min }$ holds.
The third is the local pulse with a power law decay which follows
\begin{equation}
\widetilde{I}(\tau _\theta )=I_0(1-\frac{\tau _\theta -\tau
_{\theta ,\min } }{\tau _{\theta ,\max }-\tau _{\theta ,\min
}})^\mu \qquad \qquad \qquad \qquad (\tau _{\theta ,\min }<\tau
_\theta \leq \tau _{\theta ,\max }).
\end{equation}
The peak of this intensity is at $\tau _{\theta ,\min }$. In the
case of $ \mu =2$, the relation of $\tau _{\theta ,\max }=\Delta
\tau _{\theta ,FWHM}/(1-1/\sqrt{2})+\tau _{\theta ,\min }$ holds
as well.

We assign $\tau _{\theta ,\min }=0$ and $\mu =2$ and take $\Delta
\tau _{\theta ,FWHM}=0.01$, $0.1$, $1$, $10$, and $\Gamma =10$,
$100$, $1000$, and $\alpha _0=-1$ and $\beta _0=-2.25$, to study
the width of light curves arising from these forms of local
pulses. For local pulse (4), we adopt $ \tau _{\theta ,0}=\tau
_{\theta ,\max }/2$.

We find in the $FWHM$ --- $\nu /\nu _{0,p}$ and $FWHM1/FWHM2$
--- $\nu /\nu _{0,p}$ plots (which are omitted due to the
similarity to Fig. 1) associated with local pulse (4) that a
semi-power law relationship between each of the two pulse width
quantities and frequency could also be observed for all sets of
the adopted parameters. The power law range of frequency is quite
similar to that in the case of the local Gaussian pulse. The only
significant differences are: a) the magnitude of the width of the
expected light curve is much smaller than that in the case of the
local Gaussian pulse if the local pulse width is sufficiently
large (when the local pulse width is small enough, the observed
width of the light curve would differ slightly); b) the magnitude
of the ratio of widthes of the corresponding light curve is much
larger than that in the case of the local Gaussian pulse,
regardless how large is the local pulse width. From the $FWHM$ ---
$\nu /\nu _{0,p}$ curves we find that for all the adopted values
of the Lorentz factor ($\Gamma =10$, $100$ and $1000$), $\nu
_{high}\simeq 2.4\sim 2.5\Gamma \nu _{0,p}$; and for $ \Gamma
=10$, $100$ and $1000$, $\log \nu _{high}-\log \nu _{low}\simeq
1.19\sim 1.38$, $1.19\sim 1.38$ and $1.20\sim 1.38$, respectively.
From the $FWHM1/FWHM2$ --- $\nu /\nu _{0,p}$ curves we get $\nu
_{high}\simeq 1.7\sim 2.6\Gamma \nu _{0,p}$, $1.8\sim 2.5\Gamma
\nu _{0,p}$ and $1.8\sim 2.5\Gamma \nu _{0,p}$, for $\Gamma =10$,
$100$ and $1000$, respectively, and $\log \nu _{high}-\log \nu
_{low}\simeq 1.21\sim 1.37$, $1.20\sim 1.39$ and $ 1.20\sim 1.38$,
for $\Gamma =10$, $100$ and $1000$, respectively. The values of
$(FWHM1/FWHM2)_{\min }$ and $(FWHM1/FWHM2)_{\max }$ rely only on
the local pulse width $\Delta \tau _{\theta ,FWHM}$, being
independent of the Lorentz factor. Thus, the conclusion obtained
in the case of the local Gaussian pulse holds when adopting local
pulse (4).

Adopting local pulse (5), one obtains similar results. We find
from the relationship between the width of pulses and frequency
that $\nu _{high}\simeq 2.3\sim 2.5\Gamma \nu _{0,p}$, $2.4\sim
2.5\Gamma \nu _{0,p}$ and $2.4\sim 2.5\Gamma \nu _{0,p}$, for
$\Gamma =10$, $100$ and $1000$, respectively, and $\log \nu
_{high}-\log \nu _{low}\simeq 1.19\sim 1.26$, $1.19\sim 1.28$ and
$ 1.20\sim 1.28$, for $\Gamma =10$, $100$ and $1000$,
respectively. In addition, we get from the relationship between
the the ratio of widths, $FWHM1/FWHM2$, and frequency that $\nu
_{high}\simeq 2.5\Gamma \nu _{0,p} $, $2.5\Gamma \nu _{0,p}$ and
$2.5\sim 2.6\Gamma \nu _{0,p}$, for $ \Gamma =10 $, $100$ and
$1000$, respectively, and $\log \nu _{high}-\log \nu _{low}\simeq
1.20\sim 1.26$, $1.20\sim 1.25$ and $1.20\sim 1.25$, for $ \Gamma
=10$, $100$ and $1000$, respectively. In the same way, we get
similar results when adopting local pulse (6). From the
relationship between the width of pulses and frequency we gain
$\nu _{high}\simeq 2.3\sim 2.5\Gamma \nu _{0,p}$, $2.4\sim
2.5\Gamma \nu _{0,p}$ and $2.4\sim 2.5\Gamma \nu _{0,p}$, for
$\Gamma =10$, $100$ and $1000$, respectively, and $\log \nu
_{high}-\log
\nu _{low}\simeq 1.19\sim 1.39$, $1.20\sim 1.48$ and $1.20\sim 1.47$, for $%
\Gamma =10$, $100$ and $1000$, respectively. From the relationship
between the ratio of widths and frequency we find $\nu
_{high}\simeq 2.0\sim 2.4\Gamma \nu _{0,p}$, $2.0\sim 2.4\Gamma
\nu _{0,p}$ and $2.1\sim 2.4\Gamma \nu _{0,p}$, for $\Gamma =10$,
$100$ and $1000$,
respectively, and $\log \nu _{high}-\log \nu _{low}\simeq 1.19\sim 1.49$, $%
1.19\sim 1.51$ and $1.19\sim 1.52$, for $\Gamma =10$, $100$ and
$1000$, respectively. In both cases, the values of
$(FWHM1/FWHM2)_{\min }$ and $(FWHM1/FWHM2)_{\max }$ are
independent of the Lorentz factor as well, relying only on the
local pulse width $\Delta \tau _{\theta ,FWHM}$.

We come to the conclusion that a power law relationship between
each of the two pulse width quantities and frequency could be
observed in light curves arising from different local pulse forms.
The power law range would not be significantly influenced by the
local pulse form but the magnitudes of the width and the ratio of
widthes would be obviously affected.

\section{The relationship expected for typical hard and soft bursts}

As suggested by observation, the value of $E_p$ of bright GRBs is
mainly distributed within $100\sim 600keV$ (see Preece et al.
2000). According to the above analysis, the power law range of
many bright GRBs would be found within the energy range covering
the four channels of BATSE, which was detected by many authors.
However, for this kind of burst, the power law relationship would
fail in the energy range of Swift, or there would be a turnover in
the relationship within this energy range, assuming that the
typical Band function radiation form could approximately be
applicable. Here, we make an analysis on the relationship between
the quantities discussed above in the energy range covering
channels of both BATSE and Swift for some typical GRBs. The bursts
concerned are the so-called hard and soft bursts which are defined
as the GRBs with their peak energy $E_p$ being located above and
below the second channel of BATSE, $ E_p>100keV$ and $E_p<50keV$,
respectively. According to this definition, most of bright bursts
would belong to hard bursts, and according to Strohmayer et al.
(1998), many GINGA bursts would be soft ones.

Assume that typical hard and soft bursts differ only by the
Lorentz factor of the expanding motion of the fireball surface. As
$E_p\propto \Gamma $ (see Qin 2002), taking $E_p=250keV$ as a
typical value of the peak energy for hard bursts (see Preece et
al. 2000) and assigning $\Gamma =200$ to be the Lorentz factor of
these sources, one would find the typical value of the peak energy
of a soft burst with $\Gamma =20$ to be $E_p=25keV$ which is well
within the range of soft GRBs defined above.

The energy range concerned, which covers those of BATSE and Swift,
is divided in the following eight channels: $[E_1,E_2]=[1,2]keV$
(channel A), $[2,5]keV$ (channel B), $[5,10]keV$ (channel C),
$[10,20]keV$ (channel D), $[20,50]keV$ (channel E), $[50,100]keV$
(channel F), $ [100,300]keV$ (channel G), and $[300,1000]keV$
(channel H). The last four channels are just the four channels of
BATSE.

\subsection{For various rest frame radiation forms}

Here we make the prediction on the relationship for the typical
hard and soft bursts when different rest frame radiation forms
such as the Band function spectrum, thermal synchrotron spectrum
and Comptonized spectrum are involved.

In the case of the Band function, according to Qin (2002), we
adopt the relation of $\nu _p\simeq 1.67\Gamma \nu _{0,p}$.
Applying $ E_p=250keV$ and $\Gamma =200$ we come to $\nu
_{0,p}=0.75keVh^{-1}$, which will be applied to both the typical
hard and soft bursts.

Presented in Preece et al. (2000), we find that the low energy
power law index of bright bursts is mainly distributed within
$-2\sim 0$ and the high energy power law index is distributed
mainly within $-3.5\sim -1.5$. According to Qin (2002), the
indexes are not significantly affected by the Doppler effect of
fireballs. We therefore consider indexes within these ranges.

We calculate the $FWHM$ and the ratio of the rising width $ FWHM1$
to the decaying width $FWHM2$ of the eight channels defined above
in the case of adopting the rest frame Band function spectrum with
$(\alpha _0,\beta _0)=(-1,-2.25)$ and the local Gaussian pulse
with various widths, calculated for both the typical hard ($
\Gamma =200$ and $\nu _{0,p}=0.75keVh^{-1}$) and soft ($ \Gamma
=20$ and $\nu _{0,p}=0.75keVh^{-1}$) bursts (the corresponding
table is omitted). Displayed in Figs. 4a and 4b are the
relationships between $FWHM/FWHM_E$ and $E/keV$ and between
$FWHM1/FWHM2$ and $E/keV$, respectively, in the case when the
local Gaussian pulse with $\Delta \tau _{\theta ,FWHM}=0.1$ is
adopted, where $FWHM_E$ is the width of channel E which is just
the first channel of BATSE. One finds that, under the situation
considered here, for the typical hard burst the values of
$FWHM/FWHM_E$ and $FWHM1/FWHM2$ in the first four channels (within
the Swift range) would obviously deviate from the power law curve
determined by the data of the four BATSE channels. For the typical
soft burst, the power law range is no more in the BATSE band, but
instead, it shifts to the Swift band. We find that, in the case of
the typical hard burst, the power law index deduced from the BATSE
channels would be within $-0.18$
--- $-0.09$.

Presented in Figs. 4a and 4b are also the data in both the
Beppo-SAX and Hete-II bands. One could find that the relationships
in these two bands obey the same laws implied by those in the
eight channels adopted above. (Note that, the data in the highest
energy channel of Beppo-SAX and the highest energy channel of
Hete-II are seen to be off the corresponding relationship curves
derived from the eight channels, which is due to the wider energy
ranges attached to these two channels.)

The $FWHM$ and the ratio $ FWHM1/FWHM2$ of the eight channels in
the case of the rest frame Band function spectra with $(\alpha
_0,\beta _0)=(0,-3.5)$ and $(-1.5,-2)$ respectively for both the
typical hard and soft bursts are also calculated (tables
containing the corresponding values are omitted). We find that,
for the typical hard burst, the deviation of the data of the low
energy channels of Swift from the power law relationship deduced
from the data of the four BATSE channels could be observed in the
two cases considered here. For the typical soft burst, the power
law range would be observed in the Swift band. In the case of
$(\alpha _0,\beta _0)=(0,-3.5)$, the index of the power law
relationship deduced from the four BATSE channels for the typical
hard burst ranges from $-0.48$ to $-0.27$, while in the case of
$(\alpha _0,\beta _0)=(-1.5,-2)$, the index is confined within
$-0.07$ --- $-0.03$.

Besides these rest frame spectra, several rest frame Band function
spectra with other sets of indexes are considered and they lead to
similar results (the results are omitted).

In the case of the rest frame thermal synchrotron spectrum, we
take $\nu _{0,s}=3.5\times 10^{-3}keVh^{-1}$ (see Qin 2002 Table
3). Displayed in Figs. 4c and 4d are the two relationships in the
case of adopting Gaussian pulse (3) with $\Delta \tau _{\theta
,FWHM}=0.1$ as the local pulse. It shows that, for the typical
hard burst, both the width and the ratio of the rising to the
decaying widths in the lower energy range of Swift deviate
slightly from the power law curves obtained from the data of the
BATSE channels. For the typical soft burst, the power law range
covers all the eight channels concerned, which is much different
from that of the Band function. The most remarkable result is that
both the lower and higher band platforms disappear (for both the
typical hard and soft bursts) within the concerned channels. We
find that, for the typical hard burst, the power law index deduced
from the BATSE channels would be within $-0.22$--- $ -0.12$.

In the case of the rest frame Comptonized spectrum (where we adopt
$\alpha _{0,C}=-0.6$ as well), we take $\nu _{0,C}=0.55keVh^{-1}$
(see Qin 2002 Table 2). The relationships in the case of adopting
Gaussian pulse (3) with $\Delta \tau _{\theta ,FWHM}=0.1$ as the
local pulse are presented in Figs. 4e and 4f. Shown in these
plots, the deviation mentioned above could also be observed. The
higher band platforms disappear while the lower ones remain (at
least for the typical hard burst) within the concerned channels.
For the typical soft burst, the power law range would span over
the BATSE channels as well as a few lower energy channels next to
them. We find in this situation that, for the typical hard burst,
the power law index in the BATSE channels would be within $-0.32$
--- $-0.18$, while for the typical soft burst the power law
index in the BATSE channels would be within $-0.59$
--- $-0.17$.

\subsection{When the rest frame radiation form
varies with time}

Here, we make the prediction under the assumption that the rest
frame spectrum takes a Band function form with its indexes and
peak energy decreasing with time.

In the same way, we assign $\Gamma =200$ to the typical hard burst
and $\Gamma =20$ to the soft one. Gaussian pulse (3) with $\Delta
\tau _{\theta ,FWHM}=0.1$ is taken as the local pulse, where we
once more assign $\tau _{\theta ,0}=10\sigma + \tau _{\theta ,\min
}$ and $\tau _{\theta ,\min }=0$.

Presented in Preece et al. (2000) one could find the parameters of
high time resolution spectroscopy of 156 bright GRBs. The Band
function model, the broken power law model (including the smoothly
broken power law model), and the Comptonized spectral model were
employed to fit these sources. Identifying them with the models
they were fitted, we have three classes, where the class fitted
with the Band function contains 95 bursts (sample 1), that of the
broken power law includes 55 sources, and that of the Comptonized
class has 6. We find in sample 1 that, statistically, the low and
high energy indexes $\alpha $ and $\beta $ and the peak energy
$E_{p}$ of the sources decrease with time. Even for short bursts,
this is common. Shown in Fig. 5 are the developments of the two
indexes and the peak energy for this sample, where a relative time
scale $(t-t_{\min })/(t_{\max }-t_{\min })$ is introduced to
calculate the relevant correlations. As shown in the figure, the
regression line for the low energy index is $\alpha
=-0.63-0.20(t-t_{\min })/(t_{\max }-t_{\min })$, that for the high
energy index is $\beta =-2.44-0.42(t-t_{\min })/(t_{\max }-t_{\min
})$, and that for the peak energy is
$log(E_{p}/keV)=2.46-0.16(t-t_{\min })/(t_{\max }-t_{\min })$. As
the spectrum observed is not significantly affected by the Doppler
effect of fireballs (see Qin 2002), this suggests that, the rest
frame radiation form of the sources would develop with time as
well.

We find from sample 1 that the medians of the distribution of the
uncertainty of the three parameters are $\sigma_{E_p}=33.1keV$,
$\sigma_{\alpha}=0.136$, and $\sigma_{\beta}=0.196$, respectively,
while the medians of the distribution of the deviation (in
absolute values) of the data from the regression lines deduced
above for the three parameters are $|\Delta E_p|=97.0keV$,
$|\Delta \alpha|=0.297$, and $|\Delta \beta|=0.410$, respectively.
It shows that, in terms of statistics, the measurement
uncertainties are generally less than the dispersions of data of
the three parameters. In this section, we are interested only in
the general manner of the developments of the three parameters.
Therefore, considering the development manner illustrated above
(represented by the regression lines) is enough. Thus, let us
consider a typical evolution of rest frame indexes $\alpha _0$, $
\beta _0$ and peak energy $E_{0,p}$ following $\alpha
_0=-0.63-0.20(\tau _\theta -\tau _{\theta ,1 })/(\tau _{\theta ,2
}-\tau _{\theta ,1 })$, $\beta _0=-2.44-0.42(\tau _\theta -\tau
_{\theta ,1 })/(\tau _{\theta ,2 }-\tau _{\theta ,1 })$ and
$log(E_{0,p}/keV)=-0.06-0.16(\tau _\theta -\tau _{\theta ,1
})/(\tau _{\theta ,2 }-\tau _{\theta ,1 })$, for $\tau _{\theta ,1
}\leq \tau _{\theta}\leq \tau _{\theta ,2 }$ (to deduce the last
formula, the previously adopted relation $E_p\simeq 1.67\Gamma
E_{0,p}$ is applied to the typical hard burst for which the
Lorentz factor is assumed to be $\Gamma =200$). For $\tau
_{\theta} < \tau _{\theta ,1 }$, $\alpha _0=-0.63$, $\beta
_0=-2.44$ and $log (E_{0,p}/keV)=-0.06$, while for $\tau _{\theta}
> \tau _{\theta ,2 }$, $\alpha _0=-0.83$, $\beta _0=-2.86$ and
$log (E_{0,p}/keV)=-0.22$. As mentioned above, we employ local
Gaussian pulse (3) with $\Delta \tau _{\theta ,FWHM}=0.1$ and
assign $\tau _{\theta ,0}=10\sigma +\tau _{\theta ,\min }$ and
$\tau _{\theta ,\min }=0$ to study the relationship. We adopt
$\tau _{\theta ,1}=9\sigma +\tau _{\theta ,\min }$ and $\tau
_{\theta ,2}=11\sigma +\tau _{\theta ,\min }$. Corresponding
relationships obtained in this situation are displayed in Fig. 6.
The deviation shown above could also be observed in this figure.
For the typical soft burst, the power law range shifts to the
Swift band as well. We find that the power law index in the BATSE
channels for the typical hard burst is within $-0.27$
--- $-0.08$.

It is noticed that the peaked feature suggested above does not
show. Instead, both the lower and higher band platforms shown in
Figs. 4a and 4b remain. This might be due to the small speed of
the development of the rest frame spectrum considered here (see
what discussed below).

\section{The relationship shown in individual pulses of a BATSE GRB sample}

Presented in Kocevski et al. (2003) is a sample (the KRL sample)
of FRED pulse GRBs. We consider only the first pulse of each burst
since it is this pulse that is more closely associated with the
initial condition of the event and might be less affected by
environment. In addition, we limit our study on the sources for
which the values of the peak energy are available and the
corresponding signals are obvious enough so that the pulse widthes
of at least three channels of BATSE could be well estimated. We
find 28 bursts in the KRL sample that could meet these
requirements. For these sources, the peak energy values are taken
from Mallozzi et al. (1995). To find the central values of data of
the light curve, we simply adopt equation (22) of Kocevski et al.
(2003) to fit the corresponding light curve since we find that the
form of the function could well describe the observed profile of a
FRED pulse. The pulse width in each channel of BATSE is then
estimated with the fitting parameters.

The estimated values of the $FWHM$ of the 28 GRB pulses in various
energy channels are presented in Table 3. Relationships between
the pulse width and energy for these pulses are shown in Fig. 7.
Plotted in Fig. 7 are also the limits of the corresponding power
law ranges of these pulses estimated with their peak energies
according to the relations of $log E_{low}- log E_{p} \simeq
(-1.10 \sim -1.02)$ and $log E_{high}- log E_{p} \simeq (0.157
\sim 0.177)$ which are deduced from the typical rest frame Band
function spectrum with $\alpha _0=-1$ and $\beta _0=-2.25$ (see
section 2.1), where only the largest value of $E_{low}$ and the
smallest value of $E_{high}$ associated with the provided value of
$E_{p}$ are presented.

From Fig. 7 we find:

a) a power law range could be observed in 13 sources: \#907,
\#914, \#1406, \#1733, \#1883, \#2083, \#2483, \#2665, \#2919,
\#3143, \#3954, \#4157, \#5495;

b) a lower band platform could be observed or suspected in 8
bursts: \#973, \#1773, \#1883, \#1956, \#2919, \#3143, \#4157,
\#5495;

c) a higher band platform could be observed or suspected in 6
sources: \#907, \#2083, \#2387, \#2484, \#3886, \#3892;

d) a peaked feature could be observed or suspected in 10 GRBs:
\#1467, \#2102, \#2880, \#3155, \#3870, \#3875, \#3954, \#5478,
\#5517, \#5523.

Among the 28 sources, those belonging to the
platform-power-law-platform feature group include \#907, \#914,
\#973, \#1406, \#1733, \#1883, \#1956, \#2083, \#2387, \#2484,
\#2665, \#2919, \#3143, \#3886, \#3892, \#4157, and \#5495. Those
belonging to the peaked feature class are \#1467, \#2102, \#2880,
\#3155, \#3870, \#3875, \#3954, \#5478, \#5517, and \#5523. This
suggests that the features shown in the relationship obtained from
the 27 sources (called normal bursts) are those predicted by the
Doppler effect of fireballs. The only exception is \#5541 which
shows a sinkage, instead of a peaked, feature in the relationship,
which is not a result in our analysis.

In addition, we find that, for 14 bursts (\#907, \#914, \#1406,
\#1733, \#1883, \#2083, \#2387, \#2484, \#2665, \#2919, \#3143,
\#3892, \#3954, and \#5495), the power law ranges expected from
that associated with the typical rest frame Band function spectrum
with $\alpha _0=-1$ and $\beta _0=-2.25$ and the provided value of
$E_p$ are consistent with what derived from the observational
data. For other normal bursts (there are 13), the two power law
ranges are not in agreement. Among these 13 normal bursts, the
power law range of \#3886 is in a lower energy band than its $E_p$
suggests, while for others, the power law range is in a higher
energy band than the provided value of $E_p$ confines. If the
relation (that is associated with the typical rest frame Band
function spectrum with $\alpha _0=-1$ and $\beta _0=-2.25$) used
to derive the power law range with the provided value of $E_p$ is
approximately applicable to these sources, the difference could be
explained by assuming that the peak energies of these bursts have
been less estimated. This assumption might be true since peak
energies are always measured from time-integral spectra which must
shift to a lower energy band from the hardest spectra of the
sources. Under this interpretation, only the problem of the
behavior of \#3886 is unsolved.

\section{Discussion and conclusions}

In this paper, we study in details how the pulse width $FWHM$ and
the ratio of the rising width to the decaying width $FWHM1/FWHM2$
of GRBs are related with energy under the assumption that the
sources are in the stage of fireballs which expand
relativistically.

It can be concluded from our analysis that: a) owing to the
Doppler effect of fireballs, it is common that there exists a
power law relationship between $FWHM$ and energy and between
$FWHM1/FWHM2$ and energy within a limited range of frequency; b)
the power law range and index depend strongly on the rest frame
radiation form as well as the observed peak energy (the range
could span over more than one to five orders of magnitudes of
energy for different rest frame spectra); c) the upper and lower
limits of the power law range can be determined by the observed
peak energy $E_{p}$; d) in cases when the development of the rest
frame spectrum could be ignored, a platform-power-law-platform
feature would be formed, while in cases when the rest frame
spectrum is obviously softening with time, a peaked feature would
be observed. In addition, we find that local pulse forms affect
only the magnitude of the width and the ratio of widthes.

We perform predictions on the relationships for a typical hard
burst with $\Gamma =200$ and a typical soft burst with $\Gamma
=20$. The analysis shows that, generally, for the typical hard
burst the power range would be observed in the BATSE band while
for the typical soft burst the power law range would shift to the
Swift band. In some particular cases (e.g., when the rest frame
thermal synchrotron spectrum is adopted), the power law range
could cover the BATSE as well as Swift bands for both typical
bursts.

A sample of 28 GRBs is employed to study the relationship. We find
that, except \#5541, sources of the sample either exhibit the
platform-power-law-platform feature (including 17 bursts) or show
the peaked feature (including 10 bursts). It suggests that, for
most sources of this sample, the Doppler effect of fireballs could
indeed account for the observed relationship. As for \#5541, we
wonder if other kinds of rest frame spectral evolution such as a
soft-to-hard-to-soft manner instead of the simple decreasing
pattern could lead to its specific feature (it will deserve an
investigation later). Since the peaked feature is a signature of
the development of the rest frame spectrum, we suspect that the 10
sources with the peaked feature might undergo an obvious evolution
of radiation, while for the other 17 bursts, the development, if
it exists, might be very mild.

In the above analysis, we consider the evolution of three
parameters, the lower and higher energy indexes and the peak
energy, of the rest frame Band function spectrum. We wonder what a
role each of the three factors would play in producing the peaked
feature shown above. Here we study once more the case of the
simple evolution of indexes $\alpha _0$ and $\beta _0$ and peak
frequency $\nu_{0,p}$ considered in section 2.3, but in three
different patterns. They are as follows: a) $\alpha _0=-0.5-(\tau
_\theta -\tau _{\theta ,1 })/(\tau _{\theta ,2 }-\tau _{\theta ,1
})$ for $\tau _{\theta ,1 }\leq \tau _{\theta}\leq \tau _{\theta
,2 }$, and $\alpha _0=-0.5$ for $\tau _{\theta} < \tau _{\theta ,1
}$, and $\alpha _0=-1.5$ for $\tau _{\theta} > \tau _{\theta ,2
}$, and $\beta _0=-2$ and $log \nu_{0,p}=0.1$; b) $\beta
_0=-2-(\tau _\theta -\tau _{\theta ,1 })/(\tau _{\theta ,2 }-\tau
_{\theta ,1 })$ for $\tau _{\theta ,1 }\leq \tau _{\theta}\leq
\tau _{\theta ,2 }$, and $\beta _0=-2$ for $\tau _{\theta} < \tau
_{\theta ,1 }$, and $\beta _0=-3$ for $\tau _{\theta} > \tau
_{\theta ,2 }$, and $\alpha _0=-0.5$ and $log \nu_{0,p}=0.1$; c)
$log \nu_{0,p}=0.1-(\tau _\theta -\tau _{\theta ,1 })/(\tau
_{\theta ,2 }-\tau _{\theta ,1 })$ for $\tau _{\theta ,1 }\leq
\tau _{\theta}\leq \tau _{\theta ,2 }$, and $log \nu_{0,p}=0.1$
for $\tau _{\theta} < \tau _{\theta ,1 }$, and $log \nu_{0,p}=1.1$
for $\tau _{\theta} > \tau _{\theta ,2 }$, and $\alpha _0=-0.5$
and $\beta _0=-2$. The first pattern is associated with the
evolution of the lower energy index, the second reflects nothing
but the evolution of the higher energy index, and the third
connects with the evolution of the peak energy, of the rest frame
Band function spectrum, where, for each of the three cases, the
other two parameters are fixed. In the same way we employ local
Gaussian pulse (3) with $\Delta \tau _{\theta ,FWHM}=0.1$ to study
the relationship. We adopt $\Gamma =100$, $\tau _{\theta
,1}=9\sigma +\tau _{\theta ,\min }$ and $\tau _{\theta
,2}=11\sigma +\tau _{\theta ,\min }$, and assign $\tau _{\theta
,0}=10\sigma +\tau _{\theta ,\min }$ and $\tau _{\theta ,\min
}=0$. Displayed in Fig. 8 are the corresponding $FWHM$ --- $\nu
/\nu _{0,p,max}$ and $FWHM1/FWHM2$ --- $\nu /\nu _{0,p,max}$
curves, where $\nu_{0,p,max}$ is the largest value of $\nu_{0,p}$
adopted. One finds from Fig. 8 that the peaked feature shown in
the relationship between the width and energy (see Fig. 3) is
mainly due to the evolution of the lower energy index of the rest
frame Band function spectrum, while that shown in the relationship
between the ratio of pulse widths and energy arises from the
evolution of the higher energy index. It is interesting that no
contribution from the evolution of the peak energy of the rest
frame Band function spectrum to the features could be detected
(probably the evolution of the peak energy considered here is too
mild to produce an interesting feature).

As mentioned above, it was proposed by many authors that the power
law relationship observed in GRB pulses could arise from
synchrotron radiation (see section 1). A simple synchrotron
cooling scenario is: as the electrons cool, their average energy
becomes smaller, which causes the emission peaks at lower energy
at later time (see Kazanas et al. 1998). Recently, a power law
relationship between the total isotropic energy and $E_p$ was
revealed (Lloyd et al. 2000; Amati et al. 2002). It was suggested
that this power law relation could be expected in the case of an
optically thin synchrotron shock model for a power law
distribution of electrons (see Lloyd et al. 2000). These
considerations lead to a softening picture of the rest frame
spectrum.

Does the proposal of synchrotron radiation conflict with the
effect discussed above? To find an answer to this, it might be
helpful to remind that the Doppler effect of fireballs is only a
kinetic effect while that of synchrotron radiation is a dynamic
one. Therefore, there is no confliction between the two. As
analyzed in section 2.3, a softening of the rest frame spectrum
coupling with the Doppler effect of fireballs would lead to a
peaked feature in the relationship between the pulse width and
energy if the speed of the softening is fast enough. The observed
data of our sample (see Fig. 7) show that this is indeed the case
for some events (at least for some FRED pulse GRBs).

We are wondering if the softening of the rest frame spectrum could
lead to a much different value of the power law index. We thus
analyze the power law ranges in the upper panels of Fig. 3 and
find that the index would be confined within $-0.27$ --- $-0.18$,
which is not much different from what obtained above.

We know that light curves of most bursts are complex and do not
consists of single pulses. It was pointed out that superposition
of many pulses could create the observed diversity and complexity
of GRB light curves (Fishman et al. 1994; Norris et al. 1996; Lee
et al. 2000a, 2000b). Could our analysis be applied to all light
curves observed in GRBs? The answer is no. The Doppler effect of
firteballs is associated with the angular spreading timescale
which is proportional to $1/2\Gamma^{2}$ (see, e.g., Kobayashi,
Piran, \& Sari 1997; Piran 1999; Nakar \& Piran 2002; Ryde \&
Petrosian 2002). Our model would not be applicable to light curves
of multi-pulses which are separated by timescales larger the
angular spreading timescale.

%
%

What would happen if local pulses are close enough? Let us
consider a local pulse comprising three Gaussian forms:
\begin{equation}
\widetilde{I}(\tau _\theta )= \\
I_{0,1}\exp [-(\frac{\tau _\theta -\tau _{\theta ,0,1}
}{\sigma_{1}})^2]+I_{0,2}\exp [-(\frac{\tau _\theta -\tau _{\theta
,0,2} }{\sigma_2} )^2]+I_{0,3}\exp [-(\frac{\tau _\theta -\tau
_{\theta ,0,3} }{\sigma_3} )^2] \qquad (\tau _{\theta ,\min }\leq
\tau _\theta ),
\end{equation}
where $I_{0,1}$, $I_{0,2}$, $I_{0,3}$, $\sigma_1 $, $\sigma_2 $,
$\sigma_3 $, $\tau _{\theta ,0,1}$, $\tau _{\theta ,0,2}$, $\tau
_{\theta ,0,3}$, and $\tau _{\theta ,\min }$ are constants. We
calculate light curves of (2) arising from local pulse (7) and
emitted with the typical rest frame Band function spectral form
with $\alpha _0=-1$ and $\beta_0=-2.25$, adopting $I_{0,1}=0.15$,
$I_{0,2}=0.2$, $I_{0,3}=0.04$, $\sigma_1=0.3 $, $\sigma_2=0.2 $,
$\sigma_3=0.2 $, $\tau _{\theta ,0,1}=5\sigma_{1}+\tau _{\theta
,\min }$, $\tau _{\theta ,0,2}=\tau _{\theta ,0,1}+4\sigma_{1}$,
$\tau _{\theta ,0,3}=\tau _{\theta ,0,2}+7\sigma_{2}$, $\nu
_{0,p}=0.75keVh^{-1}$, and $ \Gamma =200$, and assigning $\tau
_{\theta ,\min }=0$. Shown in Fig. 9 are the corresponding light
curves in channels A, B, C, D, E, F, G, and H, respectively, and
presented in Fig. 10 are the relationships between $FWHM/FWHM_E$
and $E/keV$ and between $FWHM1/FWHM2$ and $E/keV$ deduced from
these light curves. We find no significant difference between
these relationships and those in Fig. 4 (the data of the typical
hard burst there).

It should be pointed out that in this paper we are interested in
cases where the Doppler effect of fireballs is important and thus
we examine only FRED pulse sources. It would not be surprised if
the results are not applicable to other forms of pulses. In the
case when the mentioned effect is not at work, a power law
relationship might also exist. If so, synchrotron radiation might
be responsible to the observed relationship. This, we believe,
also deserves a detailed investigation (probably, in this case,
the pulses concerned should be non-FRED ones).

\acknowledgments

Our thanks are given to Dr. B. Paciesas for providing us the
necessary peak energy data. This work was supported by the Special
Funds for Major State Basic Research Projects (``973'') and
National Natural Science Foundation of China (No. 10273019).

\clearpage

\begin{figure}
\plotone{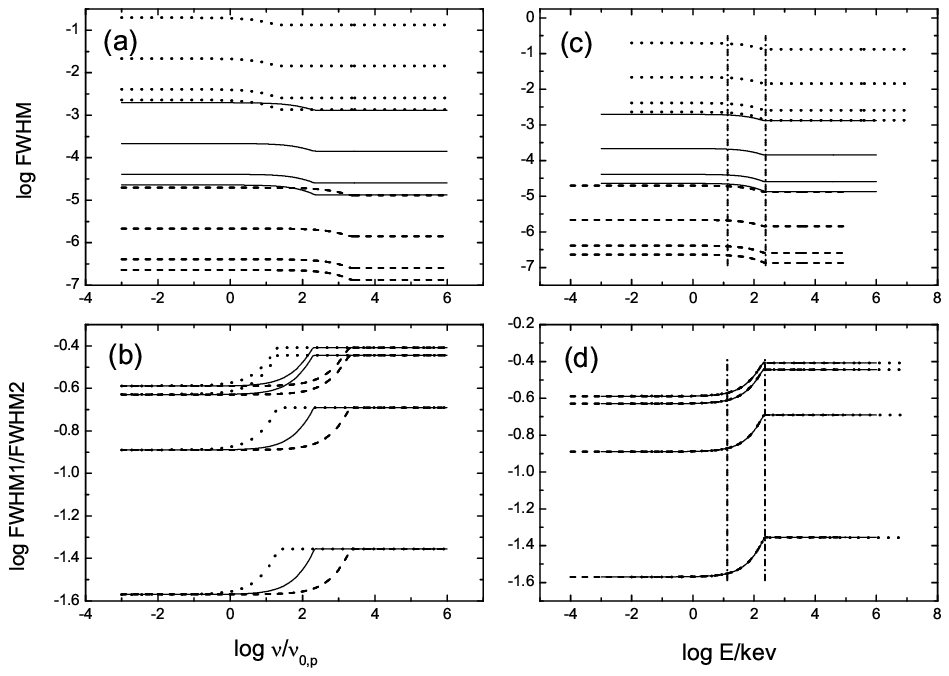}
 \caption{Relationships between
the $FWHM$ width and energy (a) and between the ratio
$FWHM1/FWHM2$ and energy (b) for the light curve of (2) confined
within $0.99\nu /\nu _{0,p}\leq \nu /\nu _{0,p}\leq 1.01\nu /\nu
_{0,p}$, in the case of adopting the Band function with $\alpha
_0=-1$\ and $\beta _0=-2.25$\ as the rest frame radiation form and
the Gaussian pulse as its local pulse. Where dot lines from the
bottom to the top represent the curves associated with $ \Delta
\tau _{\theta ,FWHM}=0.01$, $0.1$, $1$, $10$, respectively, for $
\Gamma =10$; solid lines from the bottom to the top represent the
curves associated with $\Delta \tau _{\theta ,FWHM}=0.01$, $0.1$,
$1$, $10$, respectively, for $\Gamma =100$; dash lines from the
bottom to the top stand for the curves associated with $\Delta
\tau _{\theta ,FWHM}=0.01$, $0.1$, $1$ , $10$, respectively, for
$\Gamma =1000$. Shown in panels (c) and (d) are the curves in
panels (a) and (b) respectively, where the corresponding energy is
presented in units of $keV$. The two vertical dash dot lines in
(c) and (d) denote the power law ranges deduced from the curves
associated with the case of $(\Gamma,\Delta \tau _{\theta
,FWHM})=(100,10)$.\label{Fig.1}}
\end{figure}

\clearpage

\begin{figure}
\plotone{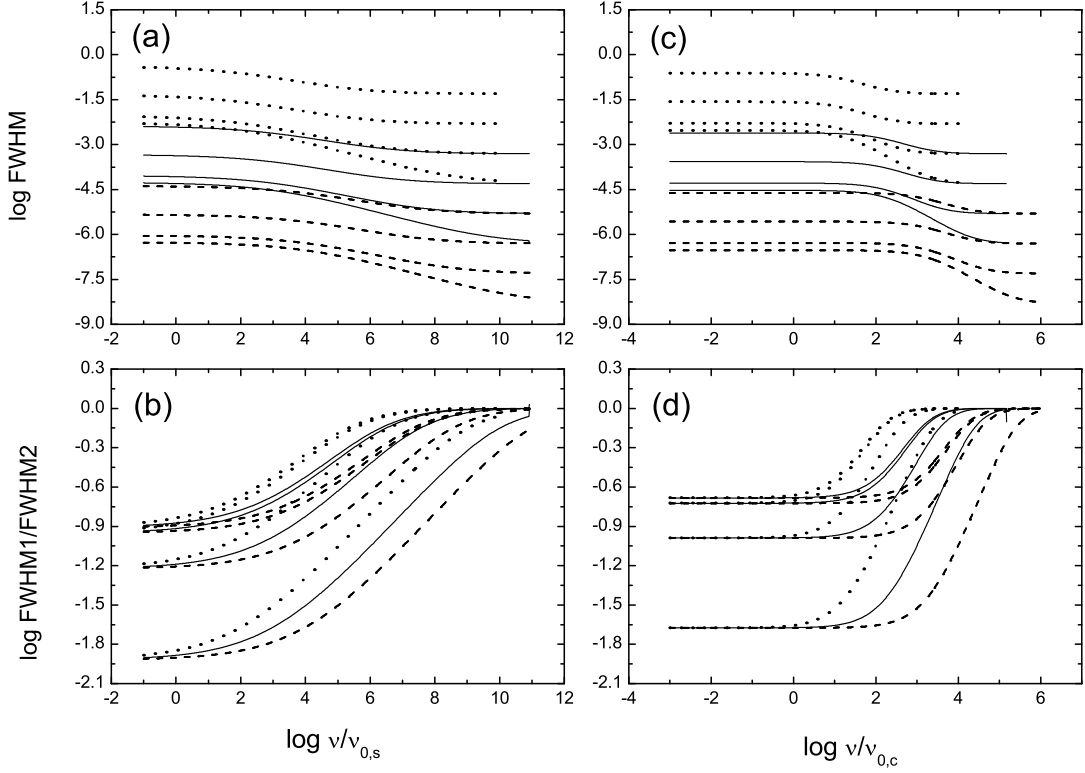}
 \caption{Relationships between
the $FWHM$, $FWHM1/FWHM2$ of pulses and energy for the light curve
of (2) confined within $0.99\nu /\nu _{0,p}\leq \nu /\nu
_{0,p}\leq 1.01\nu /\nu _{0,p}$, in the case of adopting the
thermal synchrotron spectrum (left panels) and Comptonized
spectrum (right panels) as the rest frame radiation form and the
Gaussian pulse as its local pulse. Where dot lines from the bottom
to the top represent the curves associated with $ \Delta \tau
_{\theta ,FWHM}=0.01$, $0.1$, $1$, $10$, respectively, for $
\Gamma =10$; solid lines from the bottom to the top represent the
curves associated with $\Delta \tau _{\theta ,FWHM}=0.01$, $0.1$,
$1$, $10$, respectively, for $\Gamma =100$; dash lines from the
bottom to the top stand for the curves associated with $\Delta
\tau _{\theta ,FWHM}=0.01$, $0.1$, $1$ , $10$, respectively, for
$\Gamma =1000$.  \label{Fig.2}}
\end{figure}

\clearpage

\begin{figure}
\plotone{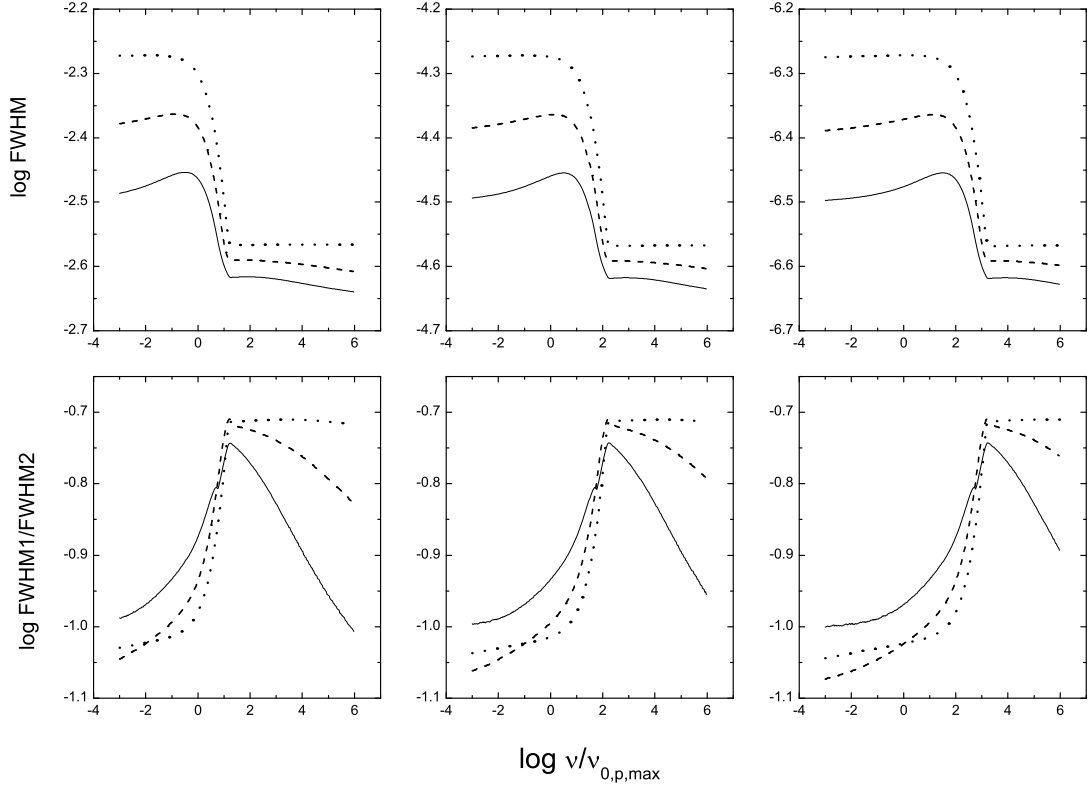} \caption{Relationship between the $FWHM$ width of
pulses and energy (upper panels) and that between $FWHM1/FWHM2$
and energy (lower panels) for the light curve of (2) confined
within $0.99\nu /\nu _{0,p}\leq \nu /\nu _{0,p}\leq 1.01\nu /\nu
_{0,p}$, in the case of adopting the varying Band function (see
section 2.3) as the rest frame radiation form and Gaussian pulse
(3) with $\Delta \tau _{\theta ,FWHM}=0.1$\ as its local pulse,
for $\Gamma =10$ (left panels), $100$ (mid panels), $1000$ (right
panels), respectively, where $\nu_{0,p,max}=10^{0.1}$. The dotted,
dashed and solid lines represent the curves with $k=0.1$, $0.5$
and $1.0$, respectively. \label{Fig. 3}}
\end{figure}

\clearpage

\begin{figure}
\plotone{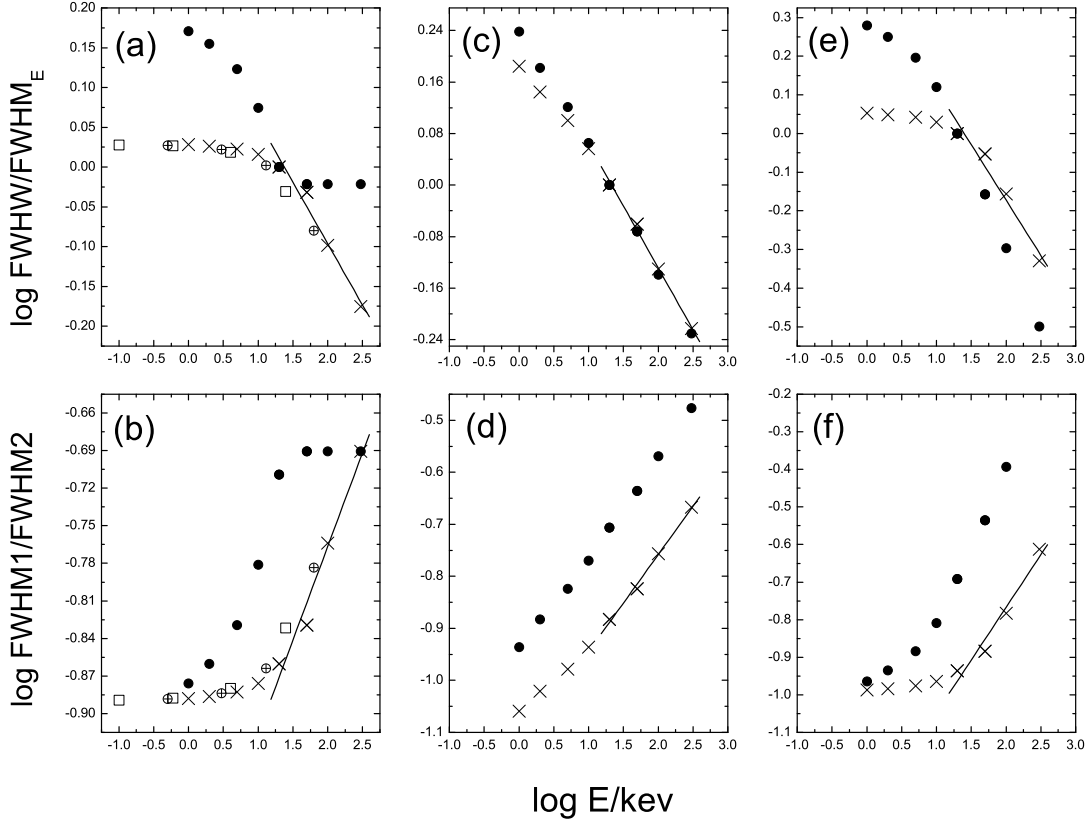} \caption{Prediction on the relationship between
the width of pulses and energy (upper panels) and that between the
ratio of the $FWHM$ width of the rising portion to that of the
decaying phase of the light curve of pulses and energy (lower
panels) for the typical hard (crosses) and soft (filled circles)
bursts. The widths are deduced from the light curve of (2)
associated with the local Gaussian pulse and the rest frame Band
function with $\alpha _0=-1$, $ \beta _0=-2.25$ and $\nu
_{0,p}=0.75keVh^{-1}$ (left panels), thermal synchrotron spectrum
with $\nu _{0,s}=3.5\times10^{-3}keVh^{-1}$ (mid panels), and
Comptonized spectrum with $\alpha _{0,C}=-0.6$, $\nu
_{0,C}=0.55keVh^{-1}$ (right panels), confined within channels A,
B, C, D, E, F, G, and H, respectively, where we adopt $\Delta\tau
_{\theta ,FWHM}=0.1$ and take $ \Gamma =200$ and $\Gamma =20$ for
typical hard and soft bursts, respectively. The solid line is the
power law curve deduced from the data of the BATSE channels for
the typical hard burst. Open squares in left panels represent the
expected data of Beppo-SAX and open circles filled with pluses
stand for those of Hete-II. \label{Fig. 4}}
\end{figure}

\clearpage

\begin{figure}
\plotone{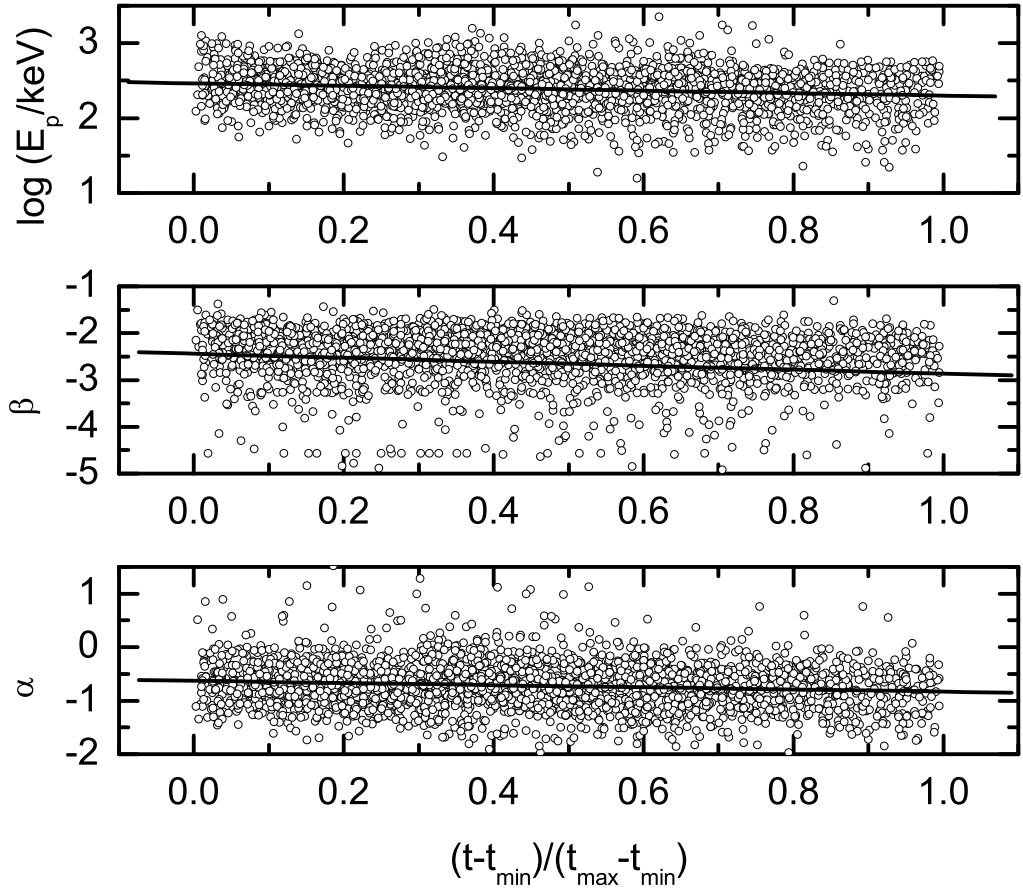} \caption{Developments of the low, high energy
indexes and the peak energy of sample 1 in terms of a relative
time scale, where $t_{min}$ and $t_{max}$ are the lower and the
upper limits of the observation time of individual sources. The
solid line is the regression line. \label{Fig. 5}}
\end{figure}

%

\clearpage

\begin{figure}
\plotone{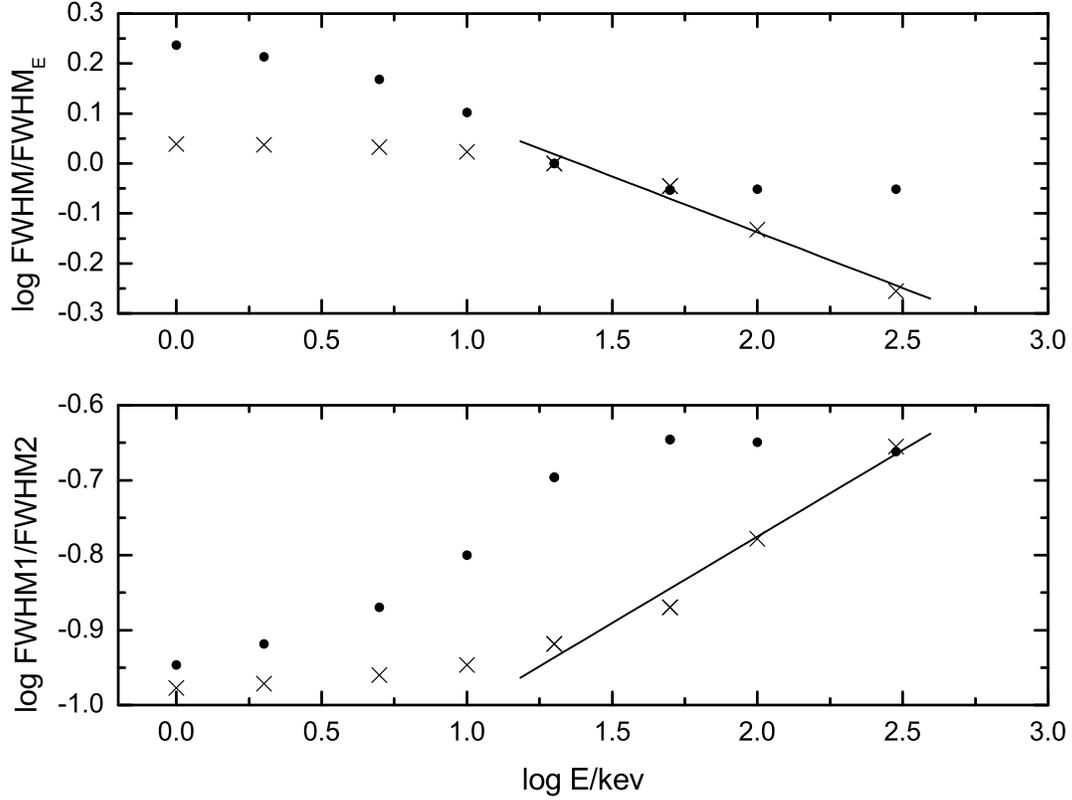} \caption{Prediction on the relationship between
the width of pulses and energy (the upper panel) and that between
the ratio of the $FWHM$ width of the rising portion to that of the
decaying phase of the light curve of pulses and energy (the lower
panel) for the typical hard (crosses) and soft (filled circles)
bursts in the case that the indexes and the peak energy of the
rest frame Band function spectrum decrease with time. The widths
are calculated from the light curve of (2) arising from local
Gaussian pulse (3), confined within channels A, B, C, D, E, F, G,
and H, respectively, where we adopt $\Delta\tau _{\theta
,FWHM}=0.1$, and we take $ \Gamma =200$ and $\Gamma =20$ for the
typical hard and soft bursts, respectively. The solid line is the
power law curve deduced from the data of the BATSE channels for
the typical hard burst. \label{Fig. 6}}
\end{figure}

\clearpage

\begin{figure}
\plotone{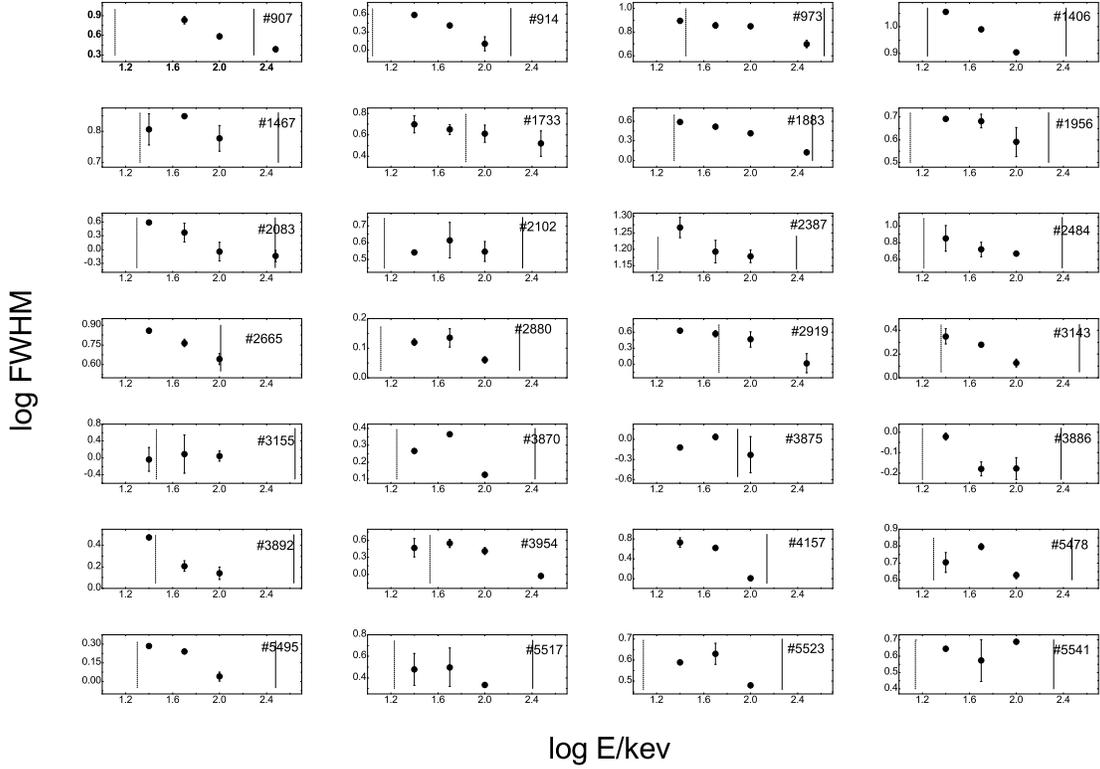} \caption{Relationship between the observed $FWHM$
width of pulses and energy shown in the BATSE energy range for the
28 GRBs concerned, where for some bursts the widths in all the 4
BATSE channels are known while for others only the widths in 3
channels are available. The dashed vertical line represents the
expected lower limit $E_{low}$ of the power law range and the
solid vertical line stands for the higher limit $E_{high}$, which
are associated with the typical rest frame Band function spectrum
with $\alpha _0=-1$ and $\beta _0=-2.25$ and the provided value of
$E_{p}$.\label{Fig.7}}
\end{figure}

\clearpage

\begin{figure}
\plotone{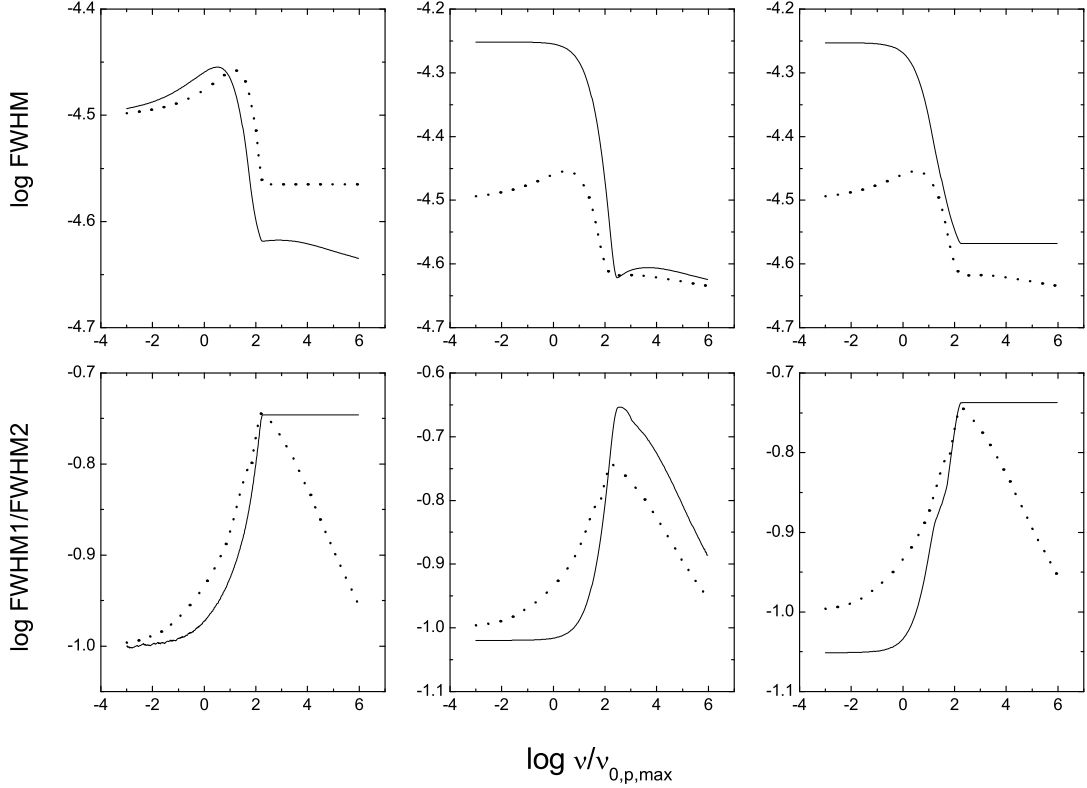} \caption{Relationship between the $FWHM$ of
pulses and energy (upper panels) and that between the ratio
$FWHM1/FWHM2$ and energy (lower panels) for the light curve of (2)
confined within $0.99\nu /\nu _{0,p}\leq \nu /\nu _{0,p}\leq
1.01\nu /\nu _{0,p}$, in the case of adopting various patterns of
development of the Band function as the rest frame radiation form
and Gaussian pulse (3) with $\Delta \tau _{\theta ,FWHM}=0.1$\ as
its local pulse and taking $\Gamma =100$ and $k=1.0$. Solid lines
in the two left panels represent the curves associated with the
case when only the lower energy index varies with time; solid
lines in the two mid panels represent those associated with the
case when only the higher energy index varies with time; solid
lines in the two right panels stand for those associated with the
case when only the peak energy varies with time. The dotted lines
represent the corresponding curves with $\Gamma =100$ and $k=1.0$
in Fig. 3 (see solid lines in the mid panels of Fig. 3).
\label{Fig. 8}}
\end{figure}

\clearpage

\begin{figure}
\plotone{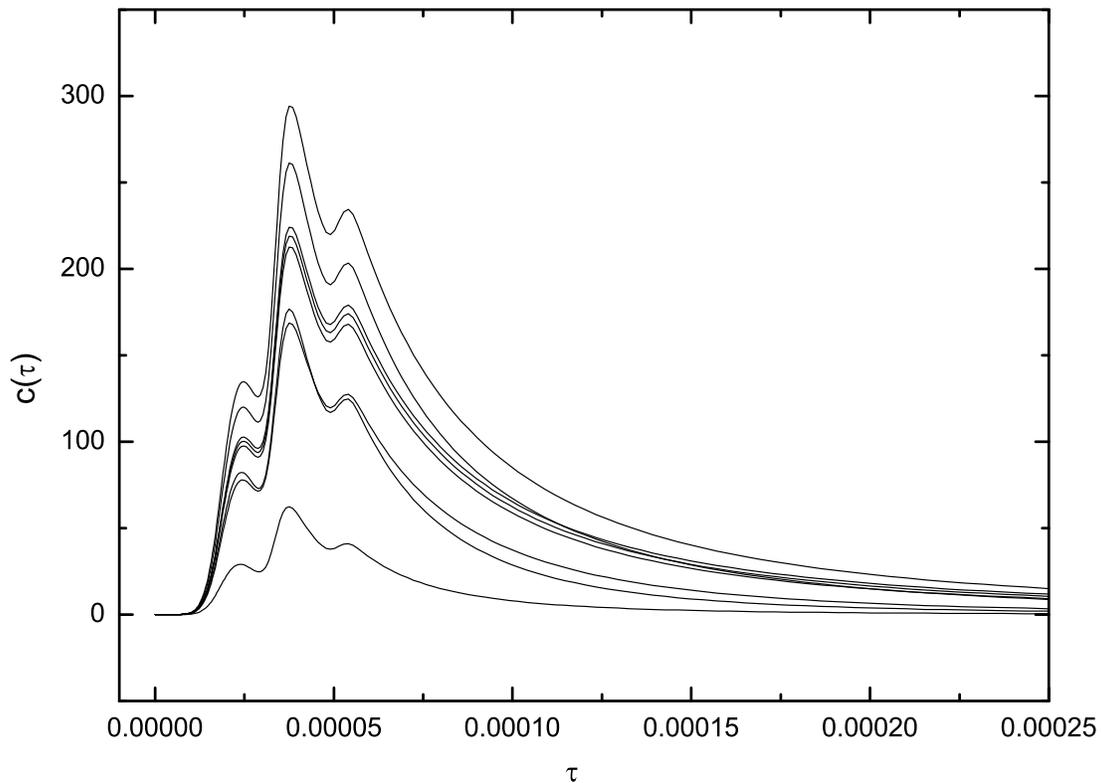} \caption{Light curves in channels B, E, A, C, D,
F, G, and H (solid lines from the top to the bottom) for the
typical hard burst ($ \Gamma =200$). The curves are calculated
with equation (2) when adopting the local pulse comprising three
Gaussian forms shown by equation (7) and the rest frame Band
function with $\alpha _0=-1$, $ \beta _0=-2.25$ and $\nu
_{0,p}=0.75keVh^{-1}$. \label{Fig. 9}}
\end{figure}

\clearpage

\begin{figure}
\plotone{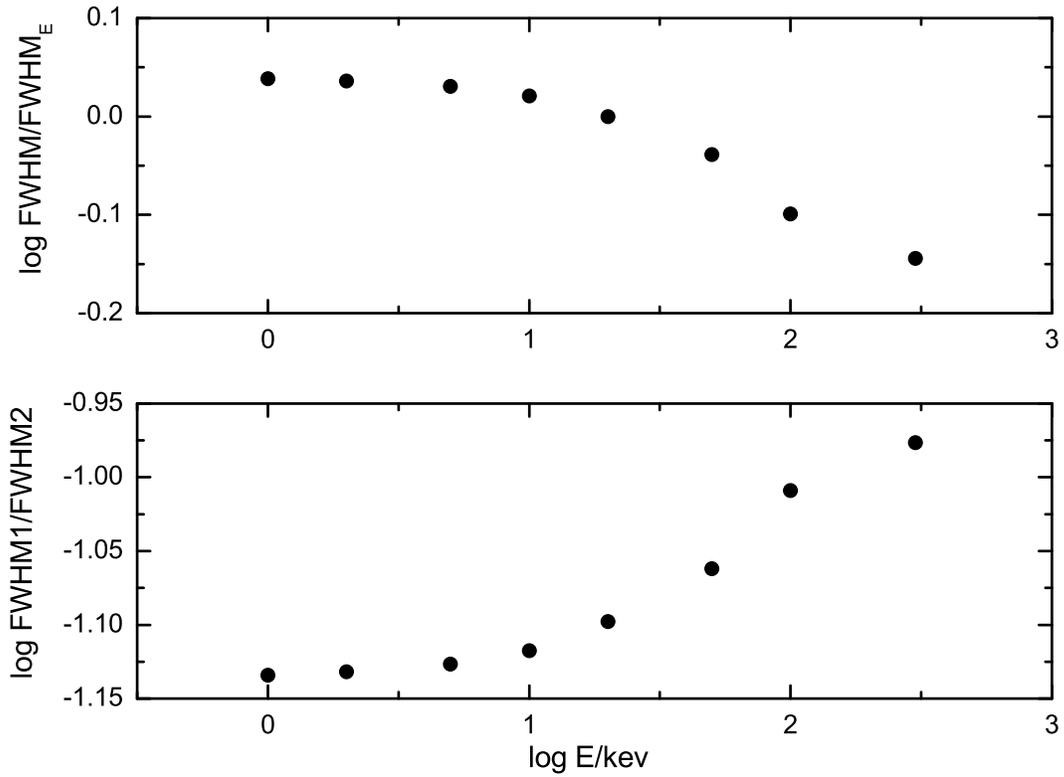} \caption{Prediction on the relationship between
the width of pulses and energy (the upper panel) and that between
the ratio of the $FWHM$ width of the rising portion to that of the
decaying phase of the light curve of pulses and energy (the lower
panel) for the typical hard burst. The widths are deduced from the
light curves of Fig. 9. \label{Fig. 10}}
\end{figure}

\clearpage

\begin{deluxetable}{llllll}
\tablecolumns{6} \tablewidth{0pc} \tablecaption{Turning frequency
and typical width obtained from the curves in Fig. 1a} \tablehead{
\colhead{$\Gamma$}   & \colhead{$\Delta\tau_{\theta,FWHM}$}    &
\colhead{log$\frac{\upsilon_{low}}{\upsilon_{0,p}}$} &
\colhead{log$\frac{\upsilon_{high}}{\upsilon_{0,p}}$}   &
\colhead{$log FWHM_{min}$}& \colhead{$log FWHM_{min}$} }
\startdata
& 0.01 & 0.20& 1.39 &-2.87& -2.64 \\
10& 0.1 & 0.17 & 1.38 &-2.59& -2.39 \\
& 1 & 0.13 & 1.38&-1.84& -1.66 \\
& 10& 0.12 & 1.38 &-0.88& -0.71\\
\hline
& 0.01 & 1.20 & 2.40 &-4.87& -4.64 \\
100& 0.1 & 1.17 & 2.40&-4.59& -4.39 \\
& 1 & 1.14& 2.38 &-3.84& -3.66  \\
& 10 & 1.13& 2.38&-2.88& -2.71\\
\hline
& 0.01 & 2.20 & 3.40 &-6.87& -6.64 \\
1000& 0.1 & 2.18& 3.39 &-6.59& -6.39\\
& 1 & 2.14 & 3.38 & -5.84& -5.66\\
& 10 & 2.13 & 3.39 & -4.88& -4.71\\
\enddata

\end{deluxetable}

\clearpage

\begin{deluxetable}{llllll}
\tablecolumns{6} \tablewidth{0pc} \tablecaption{Turning frequency
and typical width obtained from the curves in Fig. 1b} \tablehead{
\colhead{$\Gamma$}   & \colhead{$\Delta\tau_{\theta,FWHM}$}    &
\colhead{log$\frac{\upsilon_{low}}{\upsilon_{0,p}}$} &
\colhead{log$\frac{\upsilon_{high}}{\upsilon_{0,p}}$}   &
\colhead{$log(\frac{FWHM1}{FWHM2})_{min}$}&
\colhead{$log(\frac{FWHM1}{FWHM2})_{max}$} } \startdata
& 0.01 & 0.20& 1.38 &-1.57& -1.36 \\
10& 0.1 & 0.16 & 1.37 &-0.89& -0.69 \\
& 1 & 0.13 & 1.37&-0.63& -0.44 \\
& 10& 0.12 & 1.37 &-0.59& -0.41\\
\hline
& 0.01 & 1.19 & 2.39&-1.57& -1.35 \\
100& 0.1 & 1.16 & 2.36&-0.89& -0.69 \\
& 1 & 1.13& 2.36 &-0.63& -0.44 \\
& 10 & 1.12& 2.36&-0.59& -0.41\\
\hline
& 0.01 & 2.19 & 3.39 &-1.57& -1.35 \\
1000& 0.1 & 2.16& 3.36 &-0.89& -0.69\\
& 1 & 2.13& 3.35 & -0.63& -0.44\\
& 10 & 2.12 & 3.37 & -0.59& -0.41\\
\enddata

\end{deluxetable}

\clearpage

\begin{deluxetable}{lllllllll}
 \tablecolumns{9} \tablewidth{0pc} \tablecaption{Estimated values of the $FWHM$
 of the four BATSE channels for the 28 GRB sources}
 \tablehead{\colhead{trigger} &
\colhead{$W1$} & \colhead{$\sigma_{W1}$ } & \colhead{$W2$} &
\colhead{$\sigma_{W2}$ } & \colhead{$W3$} &
\colhead{$\sigma_{W3}$} & \colhead{$W4$} & \colhead{$\sigma_{W4}$}
} \startdata
907 & & &6.791& 0.898& 3.826& 0.338&2.441&0.212\\
914& 3.915& 0.176&2.562&0.159&1.269&0.350& & \\
973& 7.874& 4.530E-6 &7.177&0.393&7.072&1.818E-2&5.008& 0.360\\
1406& 11.375& 9.846E-6 &9.765& 6.605-2&8.041&5.820& &\\
1467& 6.400 &0.735 & 7.053 &1.178E-6&5.984&0.569& &\\
1733& 4.989& 0.910& 4.463 &0.454&4.069&0.746&3.308&0.896 \\
1883&3.874&0.106&3.288&0.0207&2.585&0.067&1.325&0.053\\
1956& 4.920&2.762E-2 &4.812&0.326& 3.897&0.573&&\\
2083& 3.961 & 1.675E-4& 2.365 &1.124&0.892&0.424&0.720&0.206\\
2102& 3.488 & 0.424 &4.119 &1.001& 3.526&0.477&&\\
2387& 18.447& 1.295& 15.602&1.229&15.085&0.667& & \\
2484& 7.081 & 2.503& 5.241 &1.054& 4.687& 3.200E-2 \\
2665& 7.205 & 0.312& 5.814 &0.367& 4.391&0.409& &\\
2880&1.319& 3.360E-2& 1.365 &9.737E-2&1.150&2.748E-2&& \\
2919&4.257& 8.533E-2 & 3.714&0.502& 2.906&0.942&1.045&0.429 \\
3143& 2.238& 0.329 & 1.908 &2.617& 1.334&0.101&&\\
3155&0.925 & 0.592& 1.229 &1.280 &1.044&0.302&& \\
3870& 1.842 & 2.419E-2&2.312&1.514E-2&1.336&1.264E-2&& \\
3875&0.759&1.751E-2&1.088&0.111 &0.592& 0.367&& \\
3886& 0.954 & 3.857E-2& 0.661 &5.198E-2&0.666&8.197E-2&& \\
3892& 2.979 & 2.973E-2&1.609 &0.180& 1.383&0.185&& \\
3954& 2.962 & 1.125&2.533 &0.577&2.602&0.379&0.941&0.0860\\
4157& 5.368 &1.149 & 4.209&0.470& 1.014&5.743E-2&& \\
5478& 5.052& 0.680 & 6.267 &0.251& 4.246&0.190&&\\
5495&1.929&3.197E-4&1.740&2.940E-2&1.096&9.084E-2&&\\
5517&3.004&1.032&3.146&1.296&2.156&5.173E-2&&\\
5523&3.877&5.971E-2&4.257&0.488&3.028&2.176E-2&&\\
5541&4.420&6.866E-2&3.744&1.090&4.862&5.372E-2&&\\
\enddata

Note --- $W1$, $W2$, $W3$, and $W4$ are the $FWHM$ of pulses in
the first ($20-50kev$), second ($50-100kev$), third
($100-300kev$), and fourth ($>300kev$) energy channels of BATSE,
respectively.
\end{deluxetable}


\begin{thebibliography}{Ryde and Petrosian (2002)}

\bibitem[Amati et al. (2002)]{Ama02} Amati, L., Frontera, F., Tavani, M. et al. 2002, A\&A,
390, 81

\bibitem[Band et al. (1993)]{Ban93}  Band, D., Matteson, J., Ford, L. et al.
1993, ApJ, 413, 281

\bibitem[Chiang (1998)]{Chi98}  Chiang, J., 1998, ApJ, 508, 752

\bibitem[Cohen et al. (1997)]{Coh97}  Cohen, E., Katz, J. I., Piran, T., \&
Sari, R. 1997, ApJ, 488, 330

\bibitem[Costa (1998)]{Cos98}  Costa, E. 1998, Nuclear Physics B (Proc. Suppl.), 69/1-3,
646

\bibitem[Crew et al. (2003)]{Cre03}  Crew, G. B., Lamb, D. Q., Ricker, G. R.,
Atteia, J.-L., Kawai, N., et al. 2003, ApJ, 599, 387

\bibitem[Dado et al. (2002a)]{Dad02a}  Dado, S., Dar, A., and De Rujula, A.
2002a, A\&A, 388, 1079

\bibitem[Dado et al. (2002b)]{Dad02b}  Dado, S., Dar, A., and De Rujula, A.
2002b, ApJ, 572, L143

\bibitem[Dermer (1998)]{Der98}  Dermer, C. D. 1998, ApJ, 501, L157

\bibitem[Eriksen \& Gron (2000)]{Eri00}  Eriksen, E., \& Gron, O. 2000,
Amer. J. Phys., 68, 1123

\bibitem[Fenimore et al. (1995)]{Fen95}  Fenimore, E. E., in't Zand, J. J.
M., Norris, J. P. et al. 1995, ApJ, 448, L101

\bibitem[Fenimore et al. (1996)]{Fen96}  Fenimore, E. E., Madras, C. D.,
Nayakshin, S. 1996, ApJ, 473, 998

\bibitem[Feroci et al. (2001)]{Fer01}  Feroci, M., Antonelli, L. A., Soffitta,
P., In't Zand, J. J. M., Amati, L., et al. 2001, A\&A, 378, 441

\bibitem[Fishman et al. (1992)]{Fis92}  Fishman, G., et al. 1992, in Gamma-Ray
Bursts: Huntsville, 1991, ed. W. S. Paciesas \& G. J. Fishman (New
York: AIP), 13

\bibitem[Fishman et al. (1994)]{Fis94}  Fishman, G. J., Meegan, C. A.,
Wilson, R. B. et al. 1994, ApJS, 92, 229

\bibitem[Goodman(1986)]{Goo86}  Goodman, J. 1986, ApJ, 308, L47

\bibitem[Granot et al. (1999)]{Gra99}  Granot, J., Piran, T., Sari, R. 1999,
ApJ, 513, 679

\bibitem[Hailey et al. (1999)]{Hai99}  Hailey, C. J., Harrison, F. A., Mori,
K. 1999, ApJ, 520, L25

\bibitem[Kazanas et al. (1998)]{Kaz98}  Kazanas, D., Titarchuk, L.
G., \& Hua, X.-M. 1998, ApJ, 493, 708

\bibitem[Kobayashi et al. (1997)]{Kob97} Kobayashi,
S., Piran, T., \& Sari, R. 1997, ApJ, 490, 92

\bibitem[Kocevski et al. (2003)]{Koc03}  Kocevski, D., Ryde, F., Liang, E.
2003, ApJ, 473, 998

\bibitem[Krolik \& Pier (1991)]{Kro91}  Krolik, J. H., \& Pier, E. A.
1991, ApJ, 373, 277

\bibitem[Lee et al. (2000a)]{Lee00a} Lee, A., Bloom, E. D.,
\& Petrosian, V. 2000a, ApJS, 131, 1

\bibitem[Lee et al. (2000b)]{Lee00b} Lee, A., Bloom, E. D.,
\& Petrosian, V. 2000b, ApJS, 131, 21

\bibitem[Liang et al. (1983)]{Lia83}  Liang, E. P., Jernigan, T. E.,
Rodrigues, R. 1983, ApJ, 271, 766

\bibitem[Link, Epstein, \& Priedhorsky (1993)]{Lin93}  Link, B., Epstein, R.
I., \& Priedhorsky, W. C. 1993, ApJ, 408, L81

\bibitem[Lloyd et al. (2000)]{Llo00} Lloyd, N. M., Petrosian, V., Mallozzi, R. S. 2000,
ApJ, 534, 227

\bibitem[Mallozzi et al. (1995)]{Mal95} Mallozzi, R.S., Paciesas, W.S., Pendleton, G.N., et al. 1995, Apj,
454, 597

\bibitem[Meszaros \& Rees(1998)]{Mes98}  Meszaros, P., \& Rees, M. J. 1998,
ApJ, 502, L105

\bibitem[Nakar \& Piran (2002)]{Nak02} Nakar, E., \& Piran, T. 2002, ApJ, 572, L139

\bibitem[Nemiroff (2000)]{Nem00}  Nemiroff, R. J. 2000, ApJ, 544, 805

\bibitem[Norris et al. (1996)]{Nor96}  Norris, J. P., Nemiroff, R. J.,
Bonnell, J. T. et al. 1996, ApJ, 459, 393

\bibitem[Norris et al. (2000)]{Nor00}  Norris, J. P., Marani, G. F.,
Bonnell, J. T. 2000, ApJ, 534, 248

\bibitem[Paczynski(1986)]{Pac86}  Paczynski, B. 1986, ApJ, 308, L43

\bibitem[Piran (1999)]{Pir99}  Piran, T. 1999, Phys. Rep., 314, 575

\bibitem[Piro et al. (1998)]{Pir98}  Piro, L., Heise, J., Jager, R., Costa, E.,
Frontera, F., et al. 1998, A\&A, 329, 906

\bibitem[Preece et al. (2000)]{Pre00}  Preece, R. D., Briggs, M. S.,
Mallozzi, R. S. et al. 2000, ApJS, 126, 19

\bibitem[Qin (2002)]{Qin02}  Qin, Y.-P. 2002, A\&A, 396, 705

\bibitem[Qin (2003)]{Qin03}  Qin, Y.-P. 2003, A\&A, 407, 393

\bibitem[Qin et al. (2004)]{Qin04}  Qin, Y.-P., Zhang, Z.-B., Zhang, F.-W.,
Cui, X.-H. 2004, ApJ, 617, 439 (Paper I)

\bibitem[Ryde \& Petrosian (2002)]{Ryd02}  Ryde, F., \& Petrosian, V. 2002,
ApJ, 578, 290

\bibitem[Schaefer et al. (1994)]{Sch94}  Schaefer, B. E., Teegarden, B. J.,
Fantasia, S. F., et al. 1994, ApJS, 92, 285

\bibitem[Shen et al. (2005)]{She05} Shen, R.-F., Song, L.-M., Li, Z. 2005, MNRAS, in press (astro-ph/0505276)

\bibitem[Strohmayer et al. (1998)]{Str98}  Strohmayer, T. E., Fenimore, E.
E., Murakami, T., Yoshida, A. 1998, ApJ, 500, 873

\bibitem[Wang et al. (2000)]{Wan00}  Wang, J. C., Cen, X. F., Qian, T. L., Xu,
J., \& Wang, C. Y. 2000, ApJ, 532, 267

\end{thebibliography}
\end{document}